\documentclass[12pt]{article} 
\usepackage{graphics}
\usepackage{color}
\begin{document}
\rm
\title{Dynamical features of interference phenomena in the presence of entanglement}
\author{T. Kaufherr$^{1,2,5}$, Y. Aharonov$^{1,3}$, S. Nussinov$^{1}$, S. Popescu$^{4}$, J. Tollaksen$^{3}$}
\maketitle
\noindent\emph{(1)Tel Aviv University, School of Physics and Astronomy, Tel Aviv 69978, Israel\\
(2)Physics Department, Technion-Israel Institute of Technology, Haifa 32000, Israel\\
(3)Chapman University, Schmid College of Sciences, Orange, CA 92866, USA\\
(4)H.H. Wills Physics Laboratory, University of Bristol, UK}\\
(5)E-mail: trka@post.tau.ac.il
\begin{abstract}
A "strongly" interacting, and entangling, heavy, non recoiling, external particle effects a significant change of the environment. Described locally, the corresponding entanglement event is a generalized electric Aharonov Bohm effect, that differs from the original one in a crucial way. We propose a gedanken interference experiment. The predicted shift of the interference pattern is due to a self induced or "private" potential difference experienced while the particle is in vacuum. We show that all non trivial Born Oppenheimer potentials are "private" potentials.
We apply the Born Oppenheimer approximation to interference states. Using our approach we calculate the relative phase of the external heavy particle as well as its uncertainty throughout an interference experiment /entanglement event. We thus complement the Born Oppenheimer approximation for interference states.
\end{abstract}
\section{Introduction}
In this paper we consider an "external" heavy particle that is entangled with its environment. In the past, the decoherence of such a particle was attributed to changes occurring in the environment at large. Such an explanation contradicts the principle of locality, according to which the external particle can be effected by environmental potentials at its own location only. In an article by Stern, Aharonov and Imry~\cite{stern:aharonov}, motivated by Furry and Ramsey's paper~\cite{furry:ramsey}, it was shown that decoherence can be explained locally, without taking into account changes of the environment as a whole~\cite{derived:explained}.
\newline\newline\indent
Throughout this paper we use a model where the external, heavy, non recoiling particle is initially in a superposition of two narrow wave packets $\frac{1}{\sqrt{2}}(\psi_{1}+\psi_{2})=\frac{1}{\sqrt{2}}[\psi(x)+\psi(x-L)]$, while the  environment is represented by an "internal" light particle. By "external"  we mean: accessible to direct observation or manipulation by the experimenter. "Internal" means -- not accessible to the same.
We thus discuss entanglement with the environment in terms of a specific entanglement event of the external and the internal particle. Such an entanglement event is best illustrated by a one-dimensional scattering event, where the incident light particle representing the environment is scattered from the heavy particle. Initially, the two particles are not entangled. Then, for a finite duration while the scattering goes on, the two particles are entangled. By the time the scattering is over the two particles have disentangled again. The central distinction in this paper is between "strong" and "weak" interaction. In general, the former effects a significant change of the environmental degree of freedom which is responsible for the potential in "real time", i.e., during the time of the interaction. The latter effects no such change. In the scattering example above, strong interaction means strong back-scattering of the light particle, up to total reflection. This in turn yields "strong" entanglement. By comparison,
a weakly interacting, heavy, external particle would not back-scatter the light particle. We thus consider in this paper an external, heavy, non recoiling particle  that is initially in the superposition $\Psi_{h,in}=\frac{1}{\sqrt{2}}(\psi_{1}+\psi_{2})$, strongly interacting with the environment, or light internal particle. During the ensuing entanglement event the external particle may at most pick up a relative phase between $\psi_{2}$ and $\psi_{1}$.  Thus, when the entanglement event is over, its state is given by $\Psi_{h,fin}=\frac{1}{\sqrt{2}}[\psi_{1}+e^{i\varphi_{rel}(T)}\psi_{2}]$, where $T$ is the duration of the entanglement event.
\newline\newline\indent
In this paper we ask the question: What, if it exists, is the equivalent, single (i.e., external) particle description of the strong entanglement event? This is the meaning of locality in the present context. Note that the "locality" of a quantum particle includes not only the locations of its wave packets but also the interval between them. This is demonstrated by the electric  Aharonov Bohm (AB) effect~\cite{aharonov:bohm,aharonov:kaufherr} (see below). Thus, in the local picture of the entanglement event we need only focus on the external particle's relative phase. During the strong entanglement event the external particle's instantaneous relative phase changes continuously. Starting with a definite initial value, it becomes uncertain for a finite period of time. Finally, the relative phase is again definite, but its value may have changed. Below, we show that in the local picture this evolution is due to a generalized, self induced electric AB effect.
\newline\newline\indent
Classically, there is no way to distinguish between the two force-free regions outside of an infinite, parallel-plate condenser. The reason is, that a classical particle can reside on only one side of the condenser at a time. Thus, classically, the potential difference between the plates is unobservable. By comparison, an external quantum particle   can be located on both side of the condenser at once, i.e., in a superposition $\frac{1}{\sqrt{2}}(\psi_{1}+\psi_{2})$, 1$\equiv$L, 2$\equiv$R. It follows that in quantum theory, the potential difference or, to be precise, $\int (V_{2}-V_{1})dt$ mod 2$\pi$, is observable. This is demonstrated by the electric AB effect.
\begin{figure}
\includegraphics{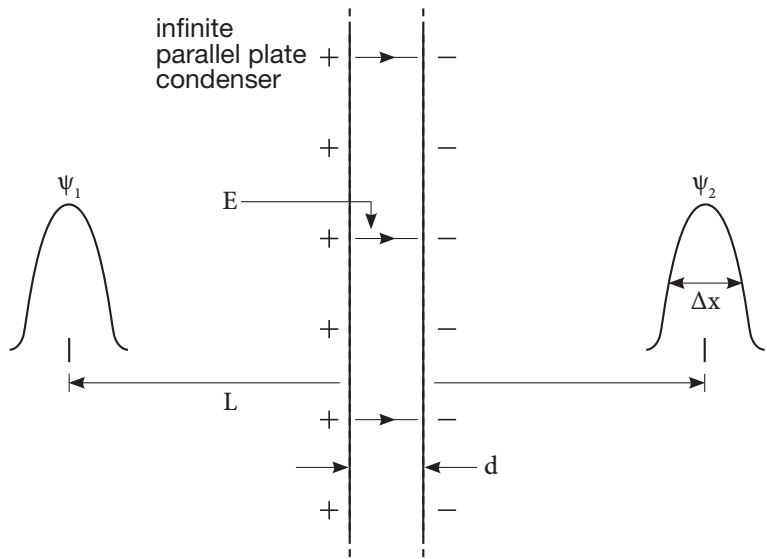}
\caption{Set-up for the electric AB effect.}
\end{figure}
In the electric AB effect an external charged particle is prepared in a wave packet $\psi$. The latter is split into two equal wave packets of width $\Delta x$ each, which are then separated by a distance $L\gg \Delta x$, and brought to rest. Thus the charged particle is initially in the state $\Psi_{in}=\frac{1}{\sqrt{2}}[\psi(x)+\psi(x-L)]\equiv\frac{1}{\sqrt{2}}(\psi_{1}+\psi_{2})$. At $t=t_{0}$, an infinite (in the $y$, $z$, directions) parallel-plate condenser is "opened" in the interval between the packets (see fig. 1 and ~\cite{finite:plates}).
The distance between the plates is $d\ll L$. At a later time $t_{1}>t_{0}$ the condenser plates are brought back together, and the electric field generated disappears. The two wave packets are then allowed to interfere.
Throughout the experiment the charged particle is confined to the force-free region outside the condenser. However, because of the electric field inside the condenser, a relative phase is generated,
\begin{eqnarray}
\Psi(t\geq t_{1})&=&\frac{1}{\sqrt{2}}[\psi_{1}+e^{i\varphi_{rel}}\psi_{2}]\,,\label{eq:I:1}\\
\varphi_{rel}(t\geq t_{1})&=&-\frac{e}{\hbar c}\int_{t_{0}}^{t_{1}}V(t)dt=-\frac{e}{\hbar c}\int_{t_{0}}^{t_{1}}\int_{0}^{d(t)}E(t)dx dt
\,,\label{eq:I:2}
\end{eqnarray}
resulting in a shift of the interference pattern. Here $V(t)=V_{2}(t)-V_{1}(t)$ is the instantaneous potential difference between the condenser's plates, $d(t)$ is their instantaneous separation and $E(t)$ is the electric field.
\newline\newline\indent
In the two-particle picture the strongly interacting non recoiling external particle entangles with the light internal particle. In the local picture the external particle experiences a corresponding, self induced, potential difference between its two wave packets. It is self induced since, if the external particle were removed, the potential would vanish. However, in quantum theory, the self induced potential has an additional, unique feature, which we call privacy. It means that the strongly interacting particle alone experiences this potential. A weakly interacting test particle, by comparison, not only would it not induce an environmental potential, but, even in the presence of a "strong" external particle, it would not monitor the private potential that the latter induces. This private potential difference that the external particle experiences, is responsible for the shift of the relative phase in the local picture.
Alternatively, the phase shift is also given by
\begin{equation}\varphi_{rel}(T)-\varphi_{rel}(0)=\int_{0}^{T}\int_{0}^{L}\langle F\rangle dxdt
\,.\label{eq:I:3}
\end{equation}
Note that the average forces $\langle F\rangle$ appearing on the right hand side (rhs) of $(\!\!~\ref{eq:I:3})$ vanish unless the external particle is actually located at the intermediate points between x=0 and x=L. Thus, with the external particle in $\psi_{1}$ and $\psi_{2}$, these forces vanish. Nevertheless, the path integral on the rhs of $(\!\!~\ref{eq:I:3})$ describes the virtual work done on the external particle while displacing it quasi statically from x=L to x=0, thus accounting for the private potential difference it experiences in the superposition.
The private potential difference depends not on the forces that exist in the interval [0,L] while the external particle is in the superposition, as is the case in the original electric AB effect, equation $(\!\!~\ref{eq:I:2})$ above. It depends, rather, on the self induced forces that would act on it at the intermediate locations.
We call the average forces "potential" forces, since when the external particle is in a superposition of $\psi_{1}$ and $\psi_{2}$ they are only a potentiality. They can be realized, but at the expense of giving up the configuration where the time integration of the private potential difference (modulo 2$\pi$) and the relative phase are observable. This is the generalized electric AB effect.
 It is, in principle, experimentally verifiable. Below (section 3), an electric AB -type gedanken interference experiment of the external particle is described. The observed shift of the interference pattern is due to a finite self induced, private potential difference experienced by the external particle while the system is in the vacuum state, i.e., force free space. To conclude: The generalized electric AB effect is contingent on our accepting the principle of locality. In the alternative, two-particle or entanglement picture, the phase shift is explained in an almost trivial way.
\newline\newline\indent
Below, the new concepts are introduced with the help of two applications. In section 2 a one-dimensional exactly solvable scattering problem is described locally. The most natural setting for the new concepts is the Born Oppenheimer approximation~\cite{born:oppenheimer}(section 3). Our discussion highlights a key aspect of the non trivial Born Oppenheimer set-ups, namely, that the external particle modifies the system with which it interacts, the non trivial Born Oppenheimer potentials being always private potentials.
We apply the Born Oppenheimer approximation to interference states. Our approach then allows us to calculate the external heavy particle's relative phase. In section 4 the uncertainty or dephasing of the relative phase in the Born Oppenheimer set-up during the interference experiment /entanglement event is calculated. We thus complement the Born Oppenheimer approximation for interference states.
\section{The Scattering Problem}
\subsection{The two-particle picture}
\begin{figure}
\includegraphics{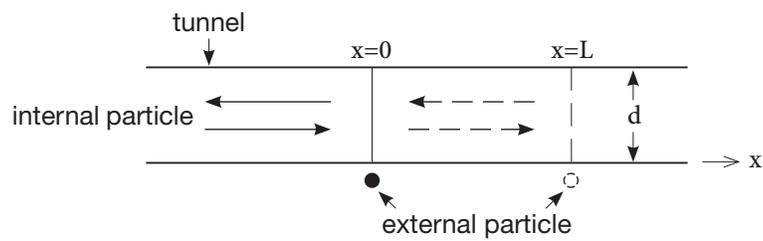}
\caption{The internal particle, moving inside the tunnel, is reflected at either
x=0 or x=L.}
\end{figure}
\begin{figure}
\includegraphics{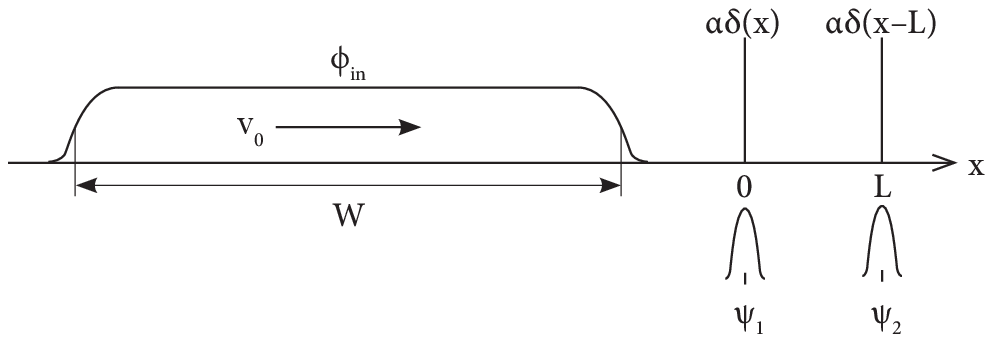}
\caption{The external particle is in a superposition of two narrow,
stationary wave packets $\psi_{1}$, $\psi_{2}$. The internal particle is initially in a very wide wave packet $\phi_{in}$, moving with velocity $v_{0}$.}
\end{figure}
Consider the following one-dimensional set-up. It consists of two particles: an "external", heavy particle, of mass M, and an "internal", light particle, representing the environment, of mass m$\ll$M. The heavy particle, h, is initially in the superposition
\begin{equation}
 \Psi_{h,in}=\frac{1}{\sqrt{2}} ( \psi_{1}+ \psi_{2})\,,
\,\label{eq:II:1}
\end{equation}
where $x_{h}$ is the position of the heavy external particle. $\psi_{1}(x_{h})$ and $\psi_{2}(x_{h})$ are two similar, narrow ($\Delta x\ll L$) wave packets, separated by a distance $L$, i.e., $\psi_{2}(x_{h})=\psi_{1}(x_{h}-L)$. The internal, light particle, l, is initially in a Gaussian wave packet $\phi_{in}(x_{l},t)$ of width $W\gg L$, moving with velocity $v_{0}$ in the positive $x$-direction inside a long, narrow tunnel (see figs. 2,3 and ~\cite{narrow:tunnel}). $x_{l}$ is the position of the light internal particle.
 The two particles interact strongly. This interaction is short-range, and we approximate it by a $\delta$ function.
$V(x_{h},x_{l})=\alpha\delta (x_{l}-x_{h})$.  For $\alpha \rightarrow\infty\;$ the light particle is  totally reflected, and complete entanglement of the two particles obtains.
We take M sufficiently large so that the recoil of the heavy particle is negligible and only momentum, but no energy is exchanged. Ignoring the constant kinetic energy of the heavy particle yields the Hamiltonian:
\begin{equation}
H=\frac{p_{l}^2}{2m}+\alpha\delta (x_{l}-x_{h})\,,
\label{eq:II:2}
\end{equation}
where $p_{l}$ is the momentum of the light particle. The state of the system at all times t is
\begin{equation}
\Psi(t)=\frac{1}{\sqrt{2}}[\psi_{1}\phi_{1}(t)+\psi_{2}\phi_{2}(t)]\,,
\label{eq:II:3}
\end{equation}
where $\psi$ and $\phi$ refer to the heavy and light particle, respectively.
Since we neglected the kinetic energy of the heavy particle, its two wave functions $\psi_{1}$ and $\psi_{2}$ are constant in time. Assuming no recoil of the heavy particle, the Hamiltonian $(\!\!~\ref{eq:II:2})$ can be approximated by (see Appendix A)
\begin{equation}
H=\frac{p_{l}^2}{2m}+\frac{1+\sigma_{3}}{2}\alpha\delta (x_{l})+\frac{1-\sigma_{3}}{2}\alpha\delta (x_{l}-L)\,,
\label{eq:II:4}
\end{equation}
where $\mid\sigma_{3}=+1>\equiv\mid\psi_{1}>$, $\mid\sigma_{3}=-1>\equiv\mid\psi_{2}>$. Within this approximation the equations for $\phi_{1}$ and $\phi_{2}$ separate, and the problem is exactly solvable for any $\alpha$ (see Appendix B).
\newline\newline\indent
The assumed large width $W$ allows us to consider both wave packets $\phi_{1}$ and $\phi_{2}$ as energy eigenstates with energy $\frac{p_{0}^{2}}{2m}$ and incident momentum $+p_{0}$. We denote by $\phi_{in}(x_{l})\approx e^{ip_{0}x_{l}}$ and by  $\phi_{out}(x_{l})\approx  e^{-ip_{0}x_{l}}$ the initial and final states of the light particle.
When $\alpha\rightarrow\infty$, the above incident state is completely reflected from either of the two $\delta$ functions. Conservation of  energy then decrees that it is reflected with momentum
$-p_{0}$ and at most a change of phase. Specifically, the reflection coefficients from $\alpha\delta(x_{l})$ and $\alpha\delta(x_{l}-L)$ for $\alpha\rightarrow\infty$ are -1 and $-e^{2ip_{0}L}$ respectively (Appendix B). For the actual wide $(W>>L)$ incident wave packet the average momentum $p_{0}$ is such that $\frac{p_{0}L}{\hbar}\gg1$ and the uncertainty $\Delta p_{l}\sim\frac{\hbar}{W}\ll\frac{\hbar}{L}$. The finite width $W$ causes a negligible uncertainty in the relative phase $\Delta\varphi_{rel}=\frac{\Delta p_{l} 2L}{\hbar}\sim \frac{2L}{W}\ll 1$. Thus, for $t>>T=\frac{W}{v_{0}}$, after the reflection is over, the system is left in a new state $\Psi_{f}=\frac{1}{\sqrt{2}}(\psi_{1}+e^{2ip_{0}L}\psi_{2})\phi_{out}$ i.e., with a shift in the relative phase between the two wave packets $\psi_{1}$ and $\psi_{2}$ of the heavy particle.
If for further simplicity we choose $2p_{0}L=\pi$, the complete wave function of the system at all times $t$ is
\begin{equation}
\Psi(t)= \sqrt{\frac{T-t}{T}}\;\frac{\psi_{1}+\psi_{2}}{\sqrt{2}}e^{ip _{0}x_{l}}+\sqrt{\frac{t}{T}}\;\frac{\psi_{1}-\psi_{2}}{\sqrt{2}}e^{-ip _{0}x_{l}}\,.
\,\label{eq:II:5}
\end{equation}
Note that for $t=0$, $\Psi(0)= \frac{\psi_{1}+\psi_{2}}{\sqrt{2}}e^{ip _{0}x_{l}}$ reproduces the initial condition  $(\!\!~\ref{eq:II:1})$.
Thus, initially, the external particle's relative phase is $\varphi_{rel}(0)=0$. By the time t=T, when the two particles have disentangled back again, it has changed to $\varphi_{rel}(T)=\pi$.
\newline In deriving equation $(\!\!~\ref{eq:II:5})$ we assumed that the reflection of a single particle in the incident wave packet of width $W$ from the two $\delta$ functions occurs instantaneously. The specific time when an individual reflection occurs is uncertain within $\Delta t=T$.
While the reflection goes on, the relative phase is uncertain, the average phase factor equals \mbox{$\left<e^{i\varphi_{rel}}(t)\right>=\frac{T-t}{T}(+1)+\frac{t}{T}(-1)\equiv P_{+}(t)(+1)+P_{-}(t)(-1)$} (see Appendix C).
This means that by the time t, the external particle has already, with probability $\frac{t}{T}$, changed its relative phase,  so that $e^{i\varphi_{rel}}=-1$. However, with probability  $\frac{T-t}{T}$, it still carries the "old" value, $e^{i\varphi_{rel}}=+1$. In general, with $\varphi_{a}, \varphi_{b}$ the initial and final relative phases, we have \mbox{$\left|\left<e^{i\varphi_{rel}}(t)\right>\right|\leq\frac{T-t}{T}\left|e^{i\varphi_{a}}\right|
+\frac{t}{T}\left|e^{i\varphi_{b}}\right|=1$}. Thus non trivial entanglement, which is the case during $0<t<T$, is characterized by the condition $\left|\left<e^{i\varphi_{rel}}(t)\right>\right|<1$.
\subsection{The local picture -- general}
We now consider the scattering problem from the local point of view. The probability for the external, heavy particle to be found at $x$ is $P_{h}(x,t)=<\Psi(t)|\delta(x_{h}-x)|\Psi(t)>$. Using equation $(\!\!~\ref{eq:II:3})$ we have,
\begin{equation}
P_{h}(x,t)=\frac{1}{2}\left[ |\psi_{1}|^{2}+|\psi_{2}|^{2}+\psi_{1}^{\ast}\psi_{2}<\phi_{1}|\phi_{2}>+cc \right]\,.
\label{eq:II:6}
\end{equation}
In the present context of entanglement, the principle of locality is the condition that all the physics must be observable through measurements of the  external particle alone, i.e., without reference to the environmental degrees of freedom. Thus, the  probability distribution $(\!\!~\ref{eq:II:6})$ must be formally equivalent to a single-particle distribution. For a single particle in the coherent  superposition  $\Psi(t)=\frac{1}{\sqrt{2}} ( \psi_{1}+e^{i\varphi_{rel}(t)} \psi_{2})$ this is
\begin{equation}
P(x,t)=\frac{1}{2}\left[ |\psi_{1}|^{2}+|\psi_{2}|^{2}+e^{i\varphi_{rel}(t)}\psi_{1}^{\ast}\psi_{2}+cc \right]\,,
\label{eq:II:7}
\end{equation}
where
\begin{equation}
\varphi_{rel}(t)=\varphi_{rel}(0)-\int_{0} ^{t}(V_{2}-V_{1})dt'\,.
\label{eq:II:8}
\end{equation}
$V_{i}$ is the potential at the location of the wave packet $\psi_{i}$. In equation $(\!\!~\ref{eq:II:7})$ the interference term is multiplied by a definite relative phase factor $e^{i\varphi_{rel}}$. But when the external particle's relative phase is uncertain as in equation $(\!\!~\ref{eq:II:5})$, we must replace the relative phase by its average, i.e., by $\left<e^{i\varphi_{rel}}\right>=\sum_{a}P_{a}e^{i\varphi_{rel,a}}$. In the above example \textit{a} obtains two values $+$ and $-$, standing for $e^{i\varphi_{rel,+}}=1$, and $e^{i\varphi_{rel,-}}=e^{2ip_{0}L}(=-1)$, respectively. Comparing equation $(\!\!~\ref{eq:II:6})$ with $(\!\!~\ref{eq:II:7})$,
the condition of locality consists of the substitution
 \begin{equation}
<\phi_{1}|\phi_{2}(t)>=\left<e^{i\varphi_{rel}}(t)\right>=\sum_{a}P_{a}e^{i\varphi_{rel,a}}
=\sum_{a}P_{a}e^{-i\int_{0}^{t}(V_{2}-V_{1})_{a}dt'}\leq 1\,.
\label{eq:II:9}
\end{equation}
The scalar product on the left hand side (lhs) depends on the external particle's position $x_{h}$ and on time. In particular, it is assumed that $V(x,t)$ is a scalar potential. $(V_{2}-V_{1})_{-}$ is the potential difference responsible for $\varphi_{rel,-}$ while $(V_{2}-V_{1})_{+}=0$.  Equation $(\!\!~\ref{eq:II:9})$ connects the two-particle or entanglement picture with the local picture. Non trivial entanglement, occurring when $<\phi_{1}|\phi_{2}><1$, which effects decoherence, is attributed in the local picture to the effect of an uncertain scalar potential difference experienced by the external particle. Finally, for t=T equation $(\!\!~\ref{eq:II:9})$ yields
 \begin{equation}
e^{i\varphi_{rel}(T)}=e^{-i\int_{0}^{T}(V_{2}-V_{1})_{b}dt}=e^{2ip_{0}L}(=-1)\,,
\label{eq:II:10}
\end{equation}
having used $<\phi_{1}|\phi_{2}(T)>=e^{2ip_{0}L}(=-1)\,,\;\;P_{+}(T)=0\,,\;\;P_{-}(T)=1$. Note that we have omitted the average sign on the lhs of $(\!\!~\ref{eq:II:10})$, since the final relative phase is definite.
\newline\newline\indent
We next proceed to calculate the potential defined by equation $(\!\!~\ref{eq:II:9})$, which is the local substitute for entanglement. As we shall see, it has highly non trivial properties.
\subsection{Two paradoxes}
We look for a solution of $(\!\!~\ref{eq:II:9})$.
Our first guess as to the potential that is responsible for the change of the external particle's relative phase leads to a paradox. With the  external particle at x=0, in $\psi_{1}$, the external particle must experience an average environmental potential~\cite{paradox:one}
\begin{equation}
\langle V\rangle(x=0)=\int P_{env}(x,t)V(x)dx=\int\phi_{1}^{\ast}\phi_{1}(x,t)\alpha\delta(x)dx\equiv \langle V\rangle_{1}
\,,\label{eq:II:11}
\end{equation}
The extremely heavy external particle then acquires an overall phase factor $e^{i\varphi_{1}(t)}=e^{-i\int^{t} \langle V\rangle_{1}dt'}$. Similarly, if the  external particle is located at x=L, in $\psi_{2}$, it acquires  an overall phase factor $e^{i\varphi_{2}(t)}=e^{-i\int^{t} \langle V\rangle_{2}dt'}$, where \mbox{$\langle V\rangle_{2}=\int\phi_{2}^{\ast}\phi_{2}(x,t)\alpha\delta(x-L)dx$}. It then follows that, with the  external particle in the superposition $\Psi=\frac{1}{\sqrt{2}}(\psi_{1}+\psi_{2})$, it acquires a relative phase \mbox{$\varphi_{rel}=\varphi_{2}-\varphi_{1}=-\int[\langle V\rangle_{2}-\langle V\rangle_{1}]dt$}. However, the potential difference $\langle V\rangle_{2}-\langle V\rangle_{1}$ vanishes identically. Since $|\phi_{2}(x_{l})>=|\phi_{1}(x_{l}-L)>$,
\begin{equation}
\langle V\rangle_{2}=<\phi_{2}(x_{l})|\alpha\delta(x_{l}-L)|\phi_{2}(x_{l})>= <\phi_{1}(x_{l}-L)|\alpha\delta(x_{l}-L)|\phi_{1}(x_{l}-L)>=\langle V\rangle_{1}
\,.\label{eq:II:12}
\end{equation}
for real $V(x)$. Thus the relative phase must also vanish, contrary to the known facts. It seems as though a scalar potential cannot properly reproduce the effects of entanglement with the environment in this case.
\begin{figure}
\includegraphics{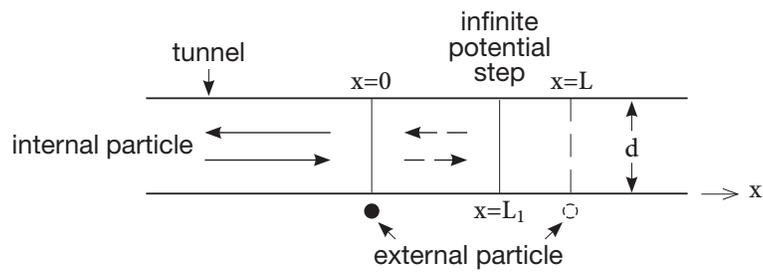}
\caption{Set-up for 2nd paradox, where an additional infinitely heavy barrier has been introduced at $L_{1}$.}
\end{figure}
\newline\newline\indent
Another paradox seems to arise if
 we add to the set-up an infinite potential step at an intermediate point $0<L_{1}< L$ (see fig. 4). A calculation using the two-particle picture tells us that in this case the external particle's relative phase changes in proportion to $L_{1}$ (Appendix D). However, if the external particle is in $\psi _{1}$, it totally reflects  the internal particle, while if it is in $\psi _{2}$, then its putative interaction with the internal particle is completely preempted by the potential step. In both cases, no information about the position $L_{1}$ of the barrier can be conveyed to the external particle by average potentials which it supposedly experiences in the local picture. How, then, can the  external particle "know" the location of the barrier?
\subsection{Resolving the first paradox -- the "private" potential}
We thus return to the defining equation $(\!\!~\ref{eq:II:9})$. Taking its time derivative, we obtain (see ~\cite{complex:conjugate} and Appendixes A and E)
\begin{eqnarray}
\frac{d}{dt}\left<e^{i\varphi_{rel}}\right>
&=&\left[< \dot{\phi}_{1}(t)\mid  \phi_{2}(t) >+< \phi_{1}(t)\mid \dot{\phi} _{2}(t) >\right]
\label{eq:II:13}\\
&=&i<\phi _{1}(t)\mid \alpha\delta(x_{l})-\alpha\delta(x_{l}-L)\mid \phi_{2}(t) >
\label{eq:II:14}\\
&=&i<\phi _{1}(t)\mid \alpha\delta(x_{l})\mid \phi_{2}(t) >=-\frac{v_{0}}{W}\left(1- e^{2ip_{0} L}\right)
\,,\label{eq:II:15}
\end{eqnarray}
where  $v_{0}$ is the velocity of the internal particle. Comparing  $(\!\!~\ref{eq:II:14}),(\!\!~\ref{eq:II:15})$ with $(\!\!~\ref{eq:I:1}),(\!\!~\ref{eq:I:2})$ we conclude that the former is  proportional to a potential difference experienced by the external particle. We call $(\!\!~\ref{eq:II:14}),(\!\!~\ref{eq:II:15})$ "private" potential, to distinguish it from the usual or "public" potential experienced by a test particle, to be discussed below. Note that the contribution of the potential at $x=L$ vanishes at the limit
$\alpha\rightarrow\infty$. At this limit there is total reflection and no transmission at $x=0$, so that $<\phi _{1}(t)\mid \alpha\delta(x_{l}-L)\mid \phi_{2}(t) >=0$.  Thus only the first term  in equation $(\!\!~\ref{eq:II:14})$, involving $\alpha\delta(x_{l})$, contributes. Note that throughout the interaction the potential is uncertain, having two values, one of which is equal to zero. Integrating $(\!\!~\ref{eq:II:15})$ yields $\left<e^{i\varphi_{rel}}(t)\right>=\frac{W-v_{0}t}{W}+ \frac{v_{0}t}{W}e^{2ip_{0} L}$. By the time $t\geq T=\frac{W}{v_{0} }$, when the internal particle has been completely reflected and the wave packet $\phi _{1}$ has moved away from the origin, the private potential $<\phi _{1}\mid\alpha\delta(x_{l})\mid\phi_{2}>= 0$, and the relative phase changes no more. It has become definite -- its value is $e^{i\varphi_{rel}}(t\geq T)=e^{2ip_{0}L}$, in agreement with the result of the calculation in the two particle picture. Thus it is the potential difference proportional to $< \dot{\phi}_{1}(t)\mid  \phi_{2}(t) >+< \phi_{1}(t)\mid \dot{\phi} _{2}(t) >$ i.e., $<\phi _{1}(t)\mid \alpha\delta(x_{l})-\alpha\delta(x_{l}-L)\mid \phi_{2}(t) >$, rather than the difference of the average potentials, that is responsible for the shift of the external particle's relative phase. This resolves the first paradox.
Note that the non vanishing $<\phi _{1}\mid\alpha\delta(x_{l})\mid\phi_{2}>=\alpha\phi_{1}^{\ast}(x=0)\phi_{2}(x=0)$ is an interference term. Thus the new potential is an interference property of the environmental degree of freedom.
This also resolves the second paradox, since in that case $\phi_{2}$ is reflected from the barrier, at L$_{1}$, so that the interference term depends on L$_{1}$. However, we shall return to the second paradox below and consider it from another point of view.
\subsection{The "public" potential}
 We distinguish between "strong" and "weak" interaction. The scattering problem we are considering is an example of an external particle strongly interacting with the environment. It interacts with the internal particle via $\alpha \delta(x_{l}-x_{h})$, which, with $ \alpha \gg1$, leads to strong reflection of the internal particle. Within this context, weak interaction is given by $\varepsilon \delta (x_{l}-x_{h})$, with $\varepsilon\ll1$. This describes the interaction of an external test particle, which allows the internal particle to continue in its motion without reflection. The Hamiltonian of an external heavy, non recoiling, test particle interacting with the internal light particle is $H_{weak}=\frac{p_{t}^{2}}{2M}+\frac{p_{l}^{2}}{2m}+\varepsilon \delta(x_{l}-x_{t})$. With the internal particle in the wave packet $\phi(x_{l},t)$ described above, the test particle effectively experiences a time dependent potential. Its  Hamiltonian is $H_{test}=\frac{p_{t}^{2}}{2M}+V(x_{t},t)$ where $V(x_{t},t)=\varepsilon \phi^{*}\phi(x_{t},t)$. At a given instant, the potential is non vanishing in the interval of width $W$, the instantaneous location of the internal wave packet. Because the interaction is weak, the internal particle passes by the test particle with velocity $v_{0}$ unperturbed. Thus, with the test particle initially in a narrow wave packet $\lambda(x_{t}-x_{0},0)$ centered at $x_{0}$, it experiences a non vanishing potential for the duration $T=\frac{W}{v_{0}}$. By the time the entire wave packet has passed by, the test particle's wave packet has acquired a phase, $\lambda(x_{t}-x_{0},t\geq T)=\lambda(x_{t}-x_{0},0)e^{-i\int_{0}^{T}Vdt}\equiv e^{-i\varphi_{0}}\lambda(x_{t}-x_{0},0)$, where $\varphi_{0}=\int_{0}^{T}Vdt=\varepsilon \phi^{*}\phi T=\varepsilon \frac{1}{W}\frac{W}{v_{0}}=\frac{\varepsilon}{v_{0}}$. We call $V(x,t)$  "public" potential since it can be measured by a test particle. The public potential is always the usual potential, i.e., the virtual work done against the source when bringing the external particle from infinity to $x$, \textit{when the source is not affected by the external particle}. This corresponds to the case of an external test particle, which, by definition, does not change the environment.
Note that, with the external, strongly interacting particle in $\psi_{1}$, the public potential experienced by the test particle is $V_{1}(x_{t},t)=\varepsilon \phi^{*}_{1}\phi_{1}(x_{t},t)$, which vanishes for $x_{t}>0$, while when it is in $\psi_{2}$, the public potential is $V_{2}(x_{t},t)=\varepsilon \phi^{*}_{2}\phi_{2}(x_{t},t)$, which vanishes for  $x_{t}>L$. Note also, that the difference between private and public potentials is a new feature, a quantum characteristic of the self induced phenomena we are considering.
\subsection{Resolving the second paradox -- the "potential" force}
We shall now show that, in the local picture, the entanglement event is a generalized electric AB effect, which resolves the second paradox. We have calculated the total change of the relative phase of the external particle during the scattering event by integrating $(\!\!~\ref{eq:II:15})$ from $t=0$ to $t=T$. Alternatively, the same change in the external particle's relative phase can be obtained as follows. Let the internal particle be initially in a very wide wave packet of width $W\gg L$ and of average momentum $p_{l,init}=p_{0}$ as above.
However, unlike the previous set-up, where the  external heavy particle was initially in a superposition of two wave packets located at $x=0$ and $x=L$, respectively, here we assume that the external particle is in a single wave packet $\psi$ centered about any point in the interval $0\leq x\leq L$. The internal  particle will then be reflected at $x$, with momentum $-p_{0}$, causing the momentum of the external particle to change by $+2p_{0}$.
With $T=\frac{W}{v_{0}}$ the time it takes the internal wave packet to be reflected at x, the average momentum transferred to the external particle after time $t$
\begin{equation}
\langle\delta p_{h}\rangle(x,t)=\langle p_{h}\rangle(x,t)-\langle p_{h}\rangle(x,0)=\left(\frac{T-t}{T}\right)0+\left(\frac{t}{T}\right)2p_{0}
=\int_{0}^{t}\langle F\rangle(x,t')dt'
\label{eq:II:16}
\end{equation}
is the weighted average, with probabilities $\frac{t}{T}$ and $\frac{T-t}{T}$ of the transfers corresponding to the reflections having occurred or not. The quantum average force which is constant both in x and in t, is therefore $\langle F\rangle=\frac{2p_{0}}{T}$. We here use the average force to describe the underlying instantaneous but uncertain scattering events which involve singular forces.
\newline\newline\indent
We recall that with the  external particle at x=L, in $\psi_{2}$, the internal  particle is reflected with a reflection coefficient $A_{L}=-e^{2ip_{0}L}$ (Appendix B). In general, with the  external particle at x$_{h}$=x, the internal  particle is reflected with a coefficient $A_{x}=-e^{2ip_{0}x}=-e^{2ip_{0}x_{h}}$, or, $\phi_{out}(x_{l},t)=-\frac{1}{\sqrt{W}}e^{2ip_{0}x_{h} }e^{-ip_{0}x_{l}}=-\frac{1}{\sqrt{W}}e^{2ip_{0}x }e^{-ip_{0}x_{l}}$. The description so far has been in the two particle picture. In the local picture, where the internal  particle's coordinate is integrated out, the same phase factor is associated with the  external particle, and we have
\begin{equation}
\psi(x_{h}-x, t\geq T)=-e^{2ip_{0}x_{h}}\psi(x_{h}-x,0) =-e^{2ip_{0}x}\psi(x_{h}-x,0)\,.
\label{eq:II:17}
\end{equation}
 We use $x_{h}$ and $x$ interchangeably since the very narrow, stationary wave packet $\psi(x_{h}-x)$ behaves like a $\delta$ function at $x$. In the local picture, this constant phase factor is attributed to the impact of a force which the environment must have exerted on the extremely heavy, non recoiling  external particle during the time interval [0, T]. We have (Appendix F)
\begin{equation}
\delta\langle p_{h}\rangle(x,t)=\langle p_{h}\rangle(x,t)-\langle p_{h}\rangle(x,0)=\frac{\partial \varphi}{\partial x}(x,t)-\frac{\partial \varphi}{\partial x}(x,0)\,.
\label{eq:II:18}
\end{equation}
From $(\!\!~\ref{eq:II:17})$ it follows that the  external particle's phase, $\varphi(x,t)$, is a continuous function of position.
We can therefore write
\begin{equation}
\varphi_{rel}(t)=\varphi(L,t)-\varphi(0,t)=\int_{0}^{L}\frac{\partial \varphi}{\partial x}(x,t)dx
\;.\label{eq:II:19}
\end{equation}
From $(\!\!~\ref{eq:II:19})$ using $(\!\!~\ref{eq:II:18})$ and $(\!\!~\ref{eq:II:16})$ we finally obtain
\begin{eqnarray}
& &\varphi_{rel}(T)-\varphi_{rel}(0)=\int_{0}^{L}\left[ \frac{\partial \varphi}{\partial x}(x,T)-\frac{\partial \varphi}{\partial x}(x,0) \right]dx \nonumber \\
&=&\int_{0}^{L}\delta\langle p_{h}\rangle(x,T)dx
=\int_{0}^{L}\int_{0}^{T}\langle F\rangle dtdx
=\int_{0}^{T}\int_{0}^{L}\langle F\rangle dxdt
\;.\label{eq:II:20}
\end{eqnarray}
where $\varphi_{rel}(T)$ is the final relative phase between $\psi_{2}$ and $\psi_{1}$. In our particular example $\varphi_{rel}(0)=0$. The path integral on the rhs of $(\!\!~\ref{eq:II:20})$ is the virtual work done on the  external particle by the  average forces $\langle F\rangle$ which act on it at the intermediate points $0\leq x\leq L$. The average force $\langle F\rangle=\frac{2p_{0}}{T}$  when integrated according to $(\!\!~\ref{eq:II:20})$ yields $\varphi_{rel}(T)=\frac{2p_{0}}{T}LT=2p_{0}L$. This is the same relative phase as obtained above, by integrating the private potential difference $(\!\!~\ref{eq:II:14})$.
We conclude that the path integral of the average forces is responsible for the private potential difference given by the rhs of $(\!\!~\ref{eq:II:14}),(\!\!~\ref{eq:II:15})$, which the external particle experiences when it is in a superposition of $\psi_{1}$ and $\psi_{2}$.
Note that \textit{the average force} $\langle F\rangle=\langle F\rangle(x,t)$ \textit{appearing in} $(\!\!~\ref{eq:II:20})$ \textit{vanishes unless the external particle is actually located at x}. In the two-particle picture we know that the force is due to the short-range strong interaction $\alpha\delta(x_{l}-x_{h})\,,\;\;\alpha\gg1\,,$ which is non vanishing only where the external particle is present. According to $(\!\!~\ref{eq:II:14},\!\!~\ref{eq:II:15})$, the external particle experiences a private potential difference which effects the change of its relative phase only when it is in the superposition $\psi_{1}$ and $\psi_{2}$; thus, in this state, the rhs of $(\!\!~\ref{eq:II:20})$ involves \emph{potential} forces which are, nevertheless, responsible for the non vanishing private potential difference.
\newline\newline\indent
The average force provides the missing link and resolves the second paradox. Specifically, the spatial part of the integral on the rhs of $(\!\!~\ref{eq:II:20})$ ends at $L_{1}$, the location of the barrier, beyond which the forces vanish. Hence the $L_{1}$ dependence. For a discussion of the uncertainty of the relative phase throughout the entanglement event see Appendix G.
\subsection{The generalized electric AB effect}
Equations $(\!\!~\ref{eq:II:14}),(\!\!~\ref{eq:II:15})$ together with equation $(\!\!~\ref{eq:II:20})$ describe a -- generalized -- electric AB effect.
The path integral on the rhs of equation $(\!\!~\ref{eq:II:20})$ is the virtual work done on the external particle when displacing it quasi statically from x=0 to x=L. It is the analog, in the present context, of the line integral
$\int_{0}^{L}Edx=V_{2}-V_{1}$ of the electric field across the plates of a condenser, which is responsible for the shift of relative phase in the usual electric AB effect. However, there is  a crucial difference.
In the usual electric AB effect, the macroscopic electric field of the condenser is independent of the external charge. In the present case, the path integral of the average forces and the definite private potential difference belong to incompatible set-ups. This is the generalized  electric AB effect, as illustrated by the scattering problem.
\newline\newline\indent
Above, we have outlined a new version of an electric AB effect, which differs from the original one in an essential way. Yet this is so only if the local approach is taken seriously. In the two-particle picture, the two scattering events, occurring at x=0 and x=L, respectively, are independent. Each produces its own position dependent phase, and they are explained locally. By comparison, in the local picture the external particle alone is considered, together with the potential differences and forces that may act upon it. In this picture we have a single event: the creation of a private potential difference between x=0 and x=L, observable only by means of an interference experiment of the external particle. In the absence of any additional agent, this potential difference must be explained in terms of the virtual work done on the  external particle. The connection between $(\!\!~\ref{eq:II:14},\!\!~\ref{eq:II:15})$ and $(\!\!~\ref{eq:II:20})$ must be made. Thus, the interpretation as an electric AB effect is forced upon us in the local picture. It also follows that it is only when locality is taken seriously, i.e., as an equally valid description of physics, that the above describes a new kind of an electric AB effect, or quantum phenomenon.
\newline\newline\indent
To conclude. We have introduced the basic concepts using the simple scattering problem as an illustration. Yet it has its limitations. Firstly, in the scattering problem there is no simple connection between the private potential difference and the path integral. Secondly, we have not been able to describe an experiment that will demonstrate the effect of the private potential in an unquestionable way. These shortcomings will be remedied in the next section, where an adiabatically changing, strongly interacting system will be described, using the Born Oppenheimer approximation.
\section{The Born Oppenheimer Approximation}
\subsection{The particle and dipole set-up}
We consider a two-level system, represented here by an electric dipole whose positive charge is fixed
at the origin. Such a system can be described by the three Pauli matrices $\sigma_{1}, \sigma_{2}, \sigma_{3}.$ Let us describe the system in the representation where $\sigma_{1}$ is diagonal. The polarization in the z direction is proportional to $\sigma_{3}$ and the energy difference between the two free levels is proportional to $\sigma_{1}$.
An external, heavy, charged particle of charge Q, moving along the positive $z$ axis interacts strongly with the dipole. The interaction is
$Q\frac{\vec{d}\cdot\hat{r}}{r^{2}}=\frac{Qe\sigma_{3}}{z^{2}}\equiv \sigma_{3}f(z)$ where we have approximated the separation
$\vec{r}\approx\vec{r}_{ext}=\vec{z}$. For a given position $z$ of the external particle, the dipole's Hamiltonian is
\begin{equation}
H_{1}=\alpha\sigma_{1}+g(t)f(z)\sigma_{3}\equiv H_{F}+V(z,t)
\,,\label{eq:III:1}
\end{equation}
with $g(t)$ a slowly varying adiabatically on- and off-switching function such that $g(-\infty)=g(+\infty)=0$. The total Hamiltonian of the system is
\begin{equation}
H=\frac{p^{2}}{2M}+H_{1}
\,\label{eq:III:2}
\end{equation}
where $p=\frac{1}{i}\frac{\partial}{\partial z}$ is the  external particle's  momentum and M is its mass. Initially, at $t=-\infty$, the dipole is closed, i.e., $g(-\infty)=0$. At that time, the dipole is in its ground state: $\phi_{g}(-\infty)=|\sigma_{1}=-1>=\mid-\alpha>$. Hence the initial state is unpolarized,  $\langle\sigma_{3}\rangle(-\infty)=0$. The ground state energy is $E_{g}(-\infty)=-\alpha$. The other eigenstate is $\phi_{exc}=|\sigma_{1}=+1>=\mid+\alpha>$ with an energy gap of $2\alpha$ between the states. The external particle is prepared to be at rest in a coherent superposition \mbox{$\Psi_{xp}=\frac{1}{\sqrt{2}}(\psi_{1}+\psi_{2})$}, $\psi_{i}=\psi[z-z(i)],$ such that $\psi_{1}$ is, effectively, outside the range of the interaction. The two wave packets are enclosed in small spherical Faraday cages centered at $z(1)$, $z(2)$, respectively, with the cage containing $\psi_{1}$ being grounded. This ensures that there is no interaction between the dipole and the external particle when in $\psi_{1}$ (see fig.5).
\begin{figure}
\includegraphics{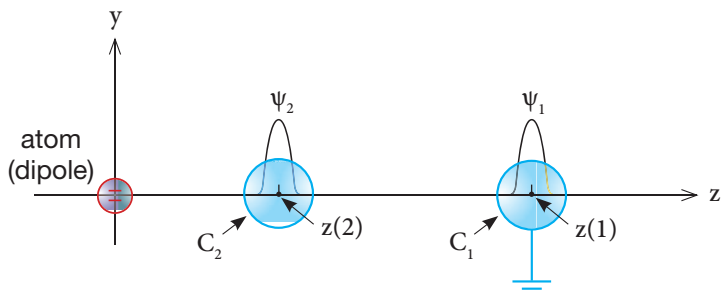}
\caption{The external particle is in a superposition of the wave packets $\psi_{1}$ and $\psi_{2}$, contained in the Faraday cages C$_{1}$, C$_{2}$ respectively.
The dipole is at the origin.}
\end{figure}
The interaction is now turned on adiabatically, until g(t) reaches at $t_{1}$ its maximal value $g(t_{1})\equiv g(0)$ such that $g(0)f[z(2)] \sim\alpha$. The adiabaticity ensures that the dipole remains in its instantaneous ground state $\phi_{g}(z,t)$ throughout, and that the  ground state energy changes according to $E_{g}(z,t)=-\sqrt{\alpha^{2}+g^{2}(t)f^{2}(z)}$ ~\cite{H1:squared}. In addition, the dipole becomes polarized (Appendix H),
\begin{equation}
\langle\sigma_{3}\rangle(z,t)=<\phi_{g}\mid\sigma_{3}\mid\phi_{g}>=\frac{-g(t)f(z)}{\sqrt{\alpha^{2}+g^{2}f^{2}}}
\,.\label{eq:III:3}
\end{equation}
When the external particle is in $\psi_{1}$ the polarization is negligible, while if it is in $\psi_{2}$, a finite dipole moment is induced. Thus the external particle and the dipole have become entangled.
We now proceed to solve for the time evolution of the  external particle.  In the Born Oppenheimer approximation~\cite{born:oppenheimer} one assumes that the light degree of freedom is in an eigenstate of energy. Thus, with the dipole prepared in the ground state of its free Hamiltonian and the adiabatic turning-on of the interaction, the Born Oppenheimer approximation is applicable. The Born Oppenheimer Hamiltonian of the  external particle  is then given by
\begin{equation}
H_{BO}=\frac{p^{2}}{2M}+E_{g}(z,t)
\,.\label{eq:III:4}
\end{equation}
The single valuedness of $E_{g}(z,t)$ is  in the adiabatic "limit". This guarantees that it is a bona-fide potential function.
 When M$\rightarrow\infty$ and the kinetic energy of the external particle is neglected, the state of the system is
\begin{equation}
\Psi(t)=\frac{1}{\sqrt{2}}\left[\psi_{1}\mid E_{1}(t)>e^{-i\int_{-\infty}^{t}E_{1}(t')dt'}
+\psi_{2}\mid E_{2}(t)>e^{-i\int_{-\infty}^{t}E_{2}(t')dt'}\right]\,,
\label{eq:III:5}
\end{equation}
where $E_{i}(t)\equiv E_{g}[z(i),t]$, $\mid E_{i}(t)>\equiv\mid E_{g}[z(i),t]>=\mid\phi_{g}[z(i),t]>$.
\subsection{The "private" and the "public" potential }
We would like now to calculate the Born Oppenheimer potential difference $ E_{g}[z(2),t]-E_{g}[z(1),t]\equiv E_{2}(t)-E_{1}(t)$ that is seen by the external particle, by integrating the forces acting on it. Consider the dipole's Hamiltonian$(\!\!~\ref{eq:III:1})$. Note that $z$, the position of the  external particle, is here a parameter. Using stationary state perturbation theory to calculate the change in the dipole's ground state energy, we obtain (Appendix I)
\begin{equation}
E_{g}[z(2),t]-E_{g}[z(1),t]=-\int_{z(1)}^{z(2)}\langle F\rangle dz
\,.\label{eq:III:6}
\end{equation}
The rhs of $(\!\!~\ref{eq:III:6})$
 is the work done when displacing the external particle quasi statically from z(1) to z(2). The average force equals $\langle F\rangle=- \langle\frac{\partial V}{\partial z}\rangle=-g(t)\langle\sigma_{3}\rangle\frac{\partial f}{\partial z}$. In general, the polarization $\langle\sigma_{3}\rangle$ may, or may not, depend on the  external particle's position. For the strongly interacting external particle we are considering, the polarization does depend on the  external particle's position and is given by $(\!\!~\ref{eq:III:3})$
above. Thus, in this case, the rhs of $(\!\!~\ref{eq:III:6})$ is a private potential difference, since it is the work done against a position dependent force. We obtain,
\begin{eqnarray}
-\int_{z(1)}^{z(2)}\langle F\rangle dz&=&g(t)\int_{z(1)}^{z(2)}\langle \sigma_{3}\rangle\frac{\partial f}{\partial z}dz
=\int_{z(1)}^{z(2)}\frac{-g^{2}(t)f(z)\frac{\partial f}{\partial z}}{\sqrt{\alpha^{2}+g^{2}f^{2}}}dz \nonumber\\
&=&\left[\sqrt{\alpha^{2}+g^{2}f^{2}}\right|_{z(1)}^{z(2)}=-\sqrt{\alpha^{2}+g^{2}f^{2}[z(2)]}+\alpha
\,,\label{eq:III:7}
\end{eqnarray}
having used $f[z(1)]=0$. Note that the rhs of $(\!\!~\ref{eq:III:7})$ is a difference of the non trivial Born Oppenheimer potential calculated earlier on, confirming $(\!\!~\ref{eq:III:6})$ for this particular case. However, it also follows from $(\!\!~\ref{eq:III:7})$ that the non trivial Born Oppenheimer potential is always a private potential. This is so since, as the calculation shows, it is the work done 'against' the position dependent polarization.
\newline\newline\indent
To put the private potential  $(\!\!~\ref{eq:III:7})$ into further perspective, let us compare it with the relevant public potential which, by definition, is obtained by integrating $(\!\!~\ref{eq:III:6})$ when the polarization is a constant. This is the case when the external particle interacts only weakly with the dipole, i.e., when it is a test particle. The Hamiltonian in this case is
\begin{eqnarray}
H_{1}'&=&\alpha\sigma_{1}+g(t)\sigma_{3}[f(z_{0})+\varepsilon  f(z_{t})]\label{eq:III:8}\\
H'_{BO}&=&\frac{p_{t}^{2}}{2M_{t}}+E_{g}'(z_{t})\,,\label{eq:III:9}
\end{eqnarray}
with $\varepsilon\ll1.$ $E_{g}'(z_{t})=E_{g}(z_{0})+\varepsilon g(t)[\langle \sigma_{3}\rangle(z_{0})]f(z_{t})$ is the ground state energy of $H_{1}'$ (Appendix J). Note that we have included a strongly interacting external particle located at a fixed position z$_{0}$, which effects a constant polarization of the atom $\langle \sigma_{3}\rangle(z_{0})$. Because of the weak coupling $\varepsilon$, \textit{the polarization does not depend on the test particle's  position}. The integration $(\!\!~\ref{eq:III:6})$ yields in this case
\begin{eqnarray}
-\int_{z(1)}^{z(2)}\langle F_{t}\rangle dz_{t}&=&g(t)[\langle \sigma_{3}\rangle(z_{0})]\int_{z(1)}^{z(2)}\frac{\partial f}{\partial z_{t}}dz_{t}\nonumber\\
&=&g(t)[\langle\sigma_{3}\rangle(z_{0})]f(z_{t})\mid_{z(1)}^{z(2)}=g(t)[\langle\sigma_{3}\rangle(z_{0})]f[z(2)]
\,,\label{eq:III:10}
\end{eqnarray}
having used $f[z(1)]=0$. This, by definition, is the public potential difference i.e., the work done on the external test particle when the environmental degree of freedom, the polarization, does not depend on the former's position. Note that $(\!\!~\ref{eq:III:10})$ is in agreement with the Born Oppenheimer potential of the test particle $E_{g}'(z_{t})$.
\subsection{The forces compared}
Note that when the external strongly interacting particle and the external test particle are located at the same point, they experience the same force $\langle F\rangle=-g(t)[\langle\sigma_{3}\rangle(z)]\frac{\partial f}{\partial z}$. We disregarded the difference in the coupling constants $\varepsilon$ versus 1.  Once the position of the "strong" particle is given, the polarization is determined. Thus, with the "strong" external particle at $z$, the force experienced by the test particle at another position, $z_{1}$ say, is $-g(t)[\langle\sigma_{3}\rangle(z)]\left(\frac{\partial f}{\partial z_{t}}\right)_{z_{1}}$. However, notwithstanding the possible difference in their magnitude, there is another, essential, difference between the forces experienced by the two particles. A test particle monitors a force that is always an independent physical reality. In our particular case the polarization is not affected by the test particle. Thus also, the force due to this polarization is an independent physical reality, as far as the test particle is concerned. Another example of such an "objective" force is the electric field inside a parallel plate capacitor. Clearly, the electric field exists independently of a test particle that may be placed inside the capacitor to measure it. By comparison, the force experienced by the strongly interacting external particle is self induced and vanishes when the external particle is removed, since then the polarization vanishes too.
\subsection{The local description of an entanglement event -- a generalized electric AB effect}
We now turn to what is the main concern of the present section which is, consideration of the particle-dipole entanglement event locally. In the final analysis, applying locality here means performing an interference experiment of the external particle. We therefore calculate the external particle's average instantaneous relative phase factor during the interference experiment /entanglement event, using $(\!\!~\ref{eq:II:3}),\;\;(\!\!~\ref{eq:III:5})$
and $(\!\!~\ref{eq:II:9})$. Writing the entangled state
as $\Psi(t)=\frac{1}{\sqrt{2}}\left[\psi_{1}\phi_{1}(t)
+\psi_{2}\phi_{2}(t)\right]$ with  $\mid \phi_{i}(t)>=\mid E_{i}(t)>e^{-i\int_{-\infty}^{t}E_{i}(t')dt'}$, we obtain
\begin{equation}
\left< e^{i\varphi_{rel}}(t)\right>=< \phi _{1}\mid \phi _{2}>=<E_{1}(t)\mid E_{2}(t)>e^{-i\int_{-\infty}^{t}[E_{2}(t')-E_{1}(t')]dt'}
\,.\label{eq:III:11}
\end{equation}
To calculate the scalar product, note that $\mid E_{1}(t)>=\mid-\alpha>$ is the ground state. $\mid E_{2}(t)>$ can be written as $\mid E_{2}(t)>=c_{1}(t)\mid E_{1}>+c_{2}(t)\mid E_{1\bot}>$, where $\mid E_{1\bot}>=\mid+\alpha>$ is orthogonal to $\mid E_{1}>$. $\mid c_{1}\mid^{2}+\mid c_{2}\mid^{2}=1$.
Thus $< E_{1}\mid E_{2}(t)>=c_{1}(t)=R(t)$. R(t) is real since the problem is two-dimensional. We obtain
\begin{equation}
\left< e^{i\varphi_{rel}}(t)\right>=R(t)e^{-i\int_{-\infty}^{t}[E_{2}(t')-E_{1}(t')]dt'}
\,.\label{eq:III:13}
\end{equation}
$\mid R(t)\mid < 1$ when the external particle and the dipole are entangled. The former's relative phase is then uncertain. This is the local manifestation of the entanglement. In general, locality means that the average instantaneous relative phase must be observable. And so it is. "Fast interference", where the  external particle's wave packets are brought together to interfere abruptly, yields the instantaneous relative phase. This is confirmed by sudden perturbation theory.
\newline\newline\indent
 The characteristic feature of an interference experiment in the present context is that, both at the beginning and at the end, the external particle and dipole are disentangled. The external particle's relative phase is thus definite at these times.  We have $\mid E_{2}(t_{initial})>=\mid E_{1}(t_{initial})>=\mid-\alpha>$ and $\mid E_{2}(t_{final})>=R(t_{final})\mid E_{1}>=R(t_{final})\mid-\alpha>$. With $R^{2}(t_{final})=1$, $R(t_{final})=\pm1$. Thus
\begin{equation}
 e^{i\varphi_{rel}}(t_{final})=e^{i\varphi_{0}}e^{-i\int_{-\infty}^{t_{final}}[E_{2}(t')-E_{1}(t')]dt'}
\,,\label{eq:III:14}
\end{equation}
where we have omitted the average sign since at $t_{final}$ the relative phase is definite. $\varphi_{0}=0\,/\pm\pi$. Note that $\varphi_{0}$ is not given by the Born Oppenheimer potentials. Investigating this additional phase and, in particular, its (possibly) more general three-dimensional counterpart, is beyond the scope of this paper. Equations $(\!\!~\ref{eq:III:13})$ and $(\!\!~\ref{eq:III:14})$ connect the particle-dipole picture with the local picture of the external particle by itself. This complements the Born Oppenheimer approximation which does not address interference states of the external particle.
Using $(\!\!~\ref{eq:III:6})$ we finally obtain for the relative phase that is  measured in the interference experiment
\begin{equation}
 \varphi_{rel}(t_{final})-\varphi_{0}=-\int_{-\infty}^{t_{final}}\left[E_{g}[z(2),t]-E_{g}[z(1),t]\right]dt
 =\int_{-\infty}^{t_{final}}\int_{z(1)}^{z(2)}\langle F\rangle dzdt
\,,\label{eq:III:15}
\end{equation}
where we have returned to the notation used in equation $(\!\!~\ref{eq:III:6})$ for the Born Oppenheimer potentials.
Thus the overall change of the external particle's relative phase during the interference experiment /entanglement event is given by the Born Oppenheimer potentials up to $\varphi_{0}=0\,/\pm\pi$. Equation $(\!\!~\ref{eq:III:15})$ describes an electric AB effect, but with a difference, since the average forces $\langle F\rangle$ replace the classical electric field of the usual electric AB effect. Note that \textit{the average force at a point vanishes unless the particle's wave packet is located at the point}. It follows that the path integral on the rhs of $(\!\!~\ref{eq:III:15})$ is the work done on the external particle when its (single) wave packet $\psi$ is actually displaced quasi statically from $z(1)$ to $z(2)$ (or from one Faraday cage to the other).
\newline\newline\indent
Given the self induced nature of the forces, this can also be understood on physical grounds. For only if the external particle is actually located in the interval between  $z(1)$ and $z(2)$, does it induce the polarization $\langle\sigma_{3}\rangle(z)$ which, in turn, gives rise to the forces $\langle F\rangle=-g(t)[\langle\sigma_{3}\rangle(z)]\frac{\partial f}{\partial z}$. Thus, when the external particle is in $\psi_{1}$ and $\psi_{2}$ the relative phase
is observable, and the forces are not observable. On the other hand, when the forces are observable, the relative phase is unobservable, since it is impossible to perform an interference experiment when the external particle is in a single wave packet. Thus, the private potential difference and the pertinent path integral of forces have become incompatible aspects of the set-up. We call these "potential" forces since, when the relative phase is observable, they are only a potentiality.
\newline\newline\indent
In the particle-dipole picture the shift of the relative phase as well as the effect of the continuum of forces are independently explained by the particle-dipole interaction. It is only in the local picture, from which the dipole is excluded, that we read the shift of the relative phase as a generalized electric AB effect. Only then is the path integral of potential forces the only possible explanation for the private potential difference, and the two "events" become interrelated. Thus, it is only when the local approach is taken seriously, that we have here by necessity a new version of an electric AB effect.
\newline\newline\indent
As is apparent from equation $(\!\!~\ref{eq:III:13})$, the Born Oppenheimer potentials,
although giving the relative phase that is measured in an interference experiment, at $t_{final}$, up to $\varphi_{0}=0\,/\pm\pi$,
are not the exact instantaneous private potentials at intermediate times, which derive from $\frac{d}{dt}<\phi_{1}\mid \phi_{2}>$.
\begin{figure}
\includegraphics{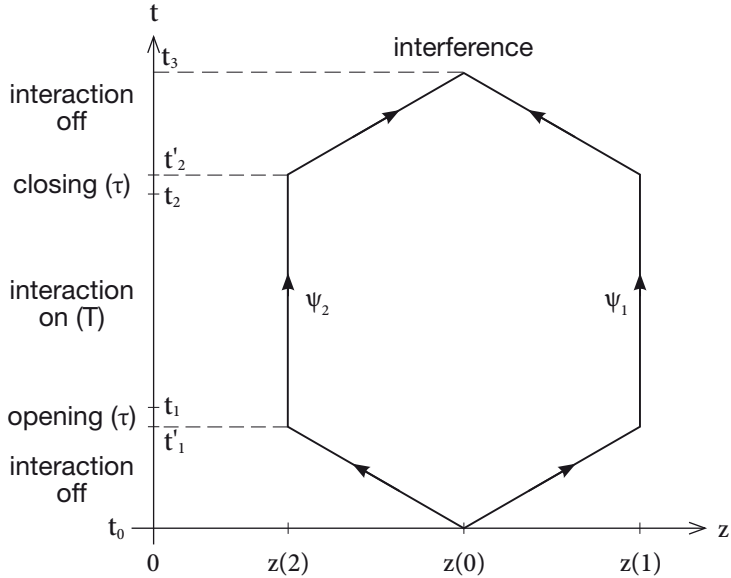}
\caption{Electric AB effect-like interference experiment.}
\end{figure}
\subsection{The gedanken experiment}
We shall next describe an interference experiment measuring the relative phase $(\!\!~\ref{eq:III:15})$ (see fig. 6). Initially, two open Faraday cages C$_{1}$ and C$_{2}$ are placed at $z(1)$, $z(2)$, respectively, and the external, heavy, strongly interacting particle is in a single wave packet $\psi$ centered at $z(0)$. At $t_{0}$ the wave packet is split into two equal parts $\psi_{1}$ and $\psi_{2}$ which are accelerated and move with  velocities  $\pm v_{t}$, respectively, until they are brought to rest inside the Faraday cages, which are then closed. The interaction is now adiabatically turned on and later again adiabatically turned off. After the interaction is turned off, when the Faraday cages are opened, the wave packets leave the cages and are brought to overlap and interfere.
The switching-on and -off function is
\begin{equation}
g(t)=\left\{ \begin{array}{ll}
                   g(0)e^{\varepsilon(t-t_{1})} &  t_{1}'\leq t\leq t_{1} \\
                   g(0) & t_{1}\leq t\leq t_{2} \\
                   g(0)e^{-\varepsilon(t-t_{2})} & t_{2}\leq t\leq t_{2}' \\
                   0 & \mbox{otherwise} \;.
                   \end{array}
            \right.
\label{eq:III:16}
\end{equation}
The interaction strength is kept at its maximal value $g(0)$ for the duration $t_{2}-t_{1}=T\gg\tau$, where $\tau$ is the finite switching-on and -off time, with $e^{-\varepsilon|\tau|}\ll 1$ guaranteeing the adiabaticity, and being sufficient to justify the formal infinite time integrations above and in section 4. The single valuedness of the adiabatic limit guarantees that the system disentangles as $g(t)$ is turned off.
Specifically,
\begin{eqnarray}
\Psi(t\geq t_{2}')=|\sigma_{1}=-1>\frac{1}{\sqrt{2}}(\psi_{1}+e^{i\varphi_{rel}(t_{2})}\psi_{2})
\,,\label{eq:III:17}\\
\varphi_{rel}(t_{2})-\varphi_{0}=-\int_{t_{1}}^{t_{2}}\left\{ E_{g}[z(2),t]-E_{g}[z(1),t] \right\} dt \label{eq:III:18}\\
=\left[\sqrt{\alpha^{2}+g^{2}(0)f^{2}[z(2)]}-\alpha\right](t_{2}-t_{1})\,,\label{eq:III:19}
\end{eqnarray}
$\varphi_{0}=0\,/\pm\pi$. We have ignored the relative phase accumulated during the opening and closing time. Repeating the experiment with N similarly prepared dipole-particle pairs, the shift of the interference pattern becomes observable. Repeating the experiment, varying the duration $T$ with the same switching-on and -off time $\tau$, equation $(\!\!~\ref{eq:III:19})$ should be confirmed.
 Note that both wave packets, being enclosed in Faraday cages while the interaction is on, are confined to force-free regions throughout the experiment.
\newline\newline\indent
We shall now show that by replacing the single dipole by two dipoles in the experiment, we can make the public potentials and the forces vanish everywhere for the duration $T$. Yet the private potential difference and the ensuing relative phase do not vanish. The Hamiltonian is
\begin{equation}
H_{1}+H_{2}=\alpha_{1}\sigma_{1}^{1}+\beta_{1} g(t)f(z)\sigma_{3}^{1}+\alpha_{2}\sigma_{1}^{2}+\beta_{2} g(t)f(z)\sigma_{3}^{2}
\,.\label{eq:III:20}
\end{equation}
 Dipole 1 is initially prepared to be in the ground state, namely the state $\mid\sigma_{1}=-1>_{1}$, while dipole 2 is prepared in the excited state, $\mid\sigma_{1}=+1>_{2}$. Thus the dipoles' polarizations are reversed.
 This implies that it is possible to make the public potential and the forces vanish everywhere during the time interval $[t_{1},t_{2}]$, by choosing $\beta_{1}\langle\sigma_{3}^{1}\rangle+\beta_{2}\langle\sigma_{3}^{2}\rangle=0$ (Appendix K).
 But there still is a non vanishing private or Born Oppenheimer potential difference between the Faraday cages (Appendix K) which will manifest in a shift of the interference pattern. Note that the condition for the vanishing of the effective total polarization is satisfied, in general, at a single point, $\tilde{z}(2)$ say. The enclosure of $\psi_{2}$ in a spherical Faraday cage centered at $\tilde{z}(2)$ guarantees that the condition is met even though the wave packet has finite width.
 \newline\newline\indent
The finite private potential difference between $\tilde{z}(2)$ and $z(1)$ seems to contradict the vanishing of the forces everywhere. In fact, there is no contradiction. The forces are monitored by test particles, when the "strong" external particle is in a superposition of $\psi_{1}$ and $\psi_{2}$, located at $z(1)$ and $\tilde{z}(2)$, respectively; while the private potential difference equals the path integral of the forces acting on the "strong" particle when it is displaced from $z(1)$ to $\tilde{z}(2)$. Note also that the total effective polarization when the external particle is at $\tilde{z}(2)$,  vanishes only during the time interval $[t_{1},t_{2}]$, or when the interaction strength is at its maximal value $g(0)$ (Appendix K). Thus, while the interaction is turned on and off, the system is not in vacuum. However, repeating the experiment varying $T$ with the same $\tau$, will prove that the observed phase shift occurred while the system was in vacuum.
\newline\newline\indent
For the sake of simplicity we have omitted in the Hamiltonian $(\!\!~\ref{eq:III:20})$
 the Coulomb interaction between the two dipoles. In the present framework this interaction, which disfavours the state when the two dipoles are parallel, is modeled by an additional interaction term $\beta\sigma_{3}^{1}\sigma_{3}^{2}$. The full Hamiltonian is $H=\alpha_{1}\sigma_{1}^{1}+\alpha_{2}\sigma_{1}^{2}+\beta\sigma_{3}^{1}\sigma_{3}^{2}
+\gamma_{1}\sigma_{3}^{1}+\gamma_{2}\sigma_{3}^{2}$. Roughly speaking, since the coupling term enhances reversed polarization, the condition for the vanishing of the public potential, i.e., $\gamma_{1}\sigma_{3}^{1}+\gamma_{2}\sigma_{3}^{2}\equiv 0$, can be readily met in this more realistic case as well. The detailed calculation may be discussed in the future.
\section{The Uncertainty of the Relative Phase in the Born Oppenheimer Approximation}
So far we mainly focused on the interference obtained after the two wave packets $\psi_{1}$ and $\psi_{2}$ were made to overlap again. In this case the Born Oppenheimer private potentials effect a shift of the relative  phase $(\!\!~\ref{eq:III:15})$,
which manifests in the interference pattern. We now address the uncertainty of the relative  phase. To this end we consider the operator (Appendix G).
\begin{equation}
\hat{\varphi}_{rel}(t)=\int_{z(1)}^{z(2)}\int_{-\infty}^{t}\hat{F}dzdt\,,
\label{eq:IV:12}
\end{equation}
from which $(\!\!~\ref{eq:III:15})$ is obtained as an expectation value at $t_{final}$. This allows us to calculate the uncertainty of the relative phase, which is the manifestation of the underlying entanglement. Specifically, we shall now show that,
\newline 1. After the dipole has been closed (at $t\rightarrow \infty$) and the external particle has disentangled from the polarizable atom, the uncertainty vanishes, i.e., $\Delta\hat{\varphi}_{rel}(\infty)=0$.
\newline 2. At intermediate times, when the external heavy particle is entangled, the uncertainty $\Delta\hat{\varphi}_{rel}(t)$ does not vanish.
\newline In the Heisenberg representation the force is given by
\begin{equation}
\hat{F}^{H}(z,t)=-\frac{\partial f}{\partial z}g(t)\sigma_{3}^{H}(t)\,.
\label{eq:IV:13}
\end{equation}
To calculate $\sigma_{3}^{H}(t)$ we write the dipole's Hamiltonian as
\begin{equation}
H_{1}=E(t)\sigma_{\theta}(t)
\,,\label{eq:IV:14}
\end{equation}
where
\begin{equation}
E(t)=\sqrt{\alpha^{2}+g^{2}(t)f^{2}(z)}\equiv \omega(t)
\,\label{eq:IV:15}
\end{equation}
is the energy (precession frequency) and
\begin{equation}
\sigma_{\theta}(t)=\cos\theta(t)\sigma_{1}+\sin\theta(t)\sigma_{3}
\,\label{eq:IV:16}
\end{equation}
is the spin operator in the direction of a fictitious magnetic field.
\begin{equation}
\cos\theta(t)=\frac{\alpha}{E(t)}\,,
\label{eq:IV:17}
\end{equation}
and
\begin{equation}
\sin\theta(t)=\frac{g(t)f(z)}{E(t)}\,.
\label{eq:IV:18}
\end{equation}
\begin{figure}
\includegraphics{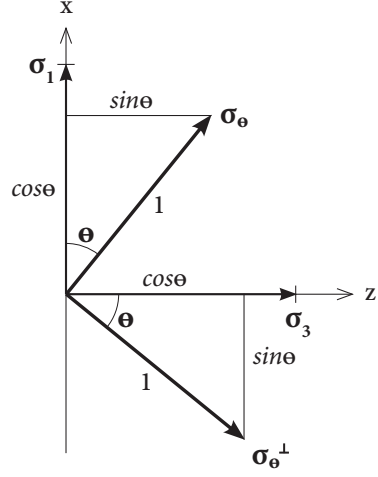}
\caption{Definition of $\theta$, $\sigma_{\theta}$, $\sigma_{\theta}^{\bot}$.}
\end{figure}
Note that $\frac{\pi}{2}-\theta$, rather than $\theta$, is in fact the usual polar angle (see also fig. 7).
The orthogonal (in the $x-z$ plane) spin operator is defined to be
\begin{equation}
\sigma_{\theta}^{\perp}(t)=-\sin\theta(t)\sigma_{1}+\cos\theta(t)\sigma_{3}
\,.\label{eq:IV:19}
\end{equation}
Using $(\!\!~\ref{eq:IV:16})$ and $(\!\!~\ref{eq:IV:19})$ we can now write
$\sigma_{3}$ in terms of $\sigma_{\theta}$ and $\sigma_{\theta}^{\perp}$. But the solution for $\sigma_{3}$ is particularly simple in the Heisenberg representation. We therefore write
\begin{equation}
\sigma_{3}^{H}(t)=\sin\theta(t)\sigma_{\theta}^{H}+\cos\theta(t)\sigma_{\theta}^{\perp H}
\,,\label{eq:IV:20}
\end{equation}
where $\sigma^{H}=U\sigma U^{\dag}$. Equation $(\!\!~\ref{eq:IV:20})$ is obtained from $(\!\!~\ref{eq:IV:16})$ and $(\!\!~\ref{eq:IV:19})$ after transforming to the Heisenberg representation.
At $t=-\infty$, $g(-\infty)=0$, $\cos\theta(-\infty)=1$ and $\sin\theta(-\infty)=0$. Thus  $\sigma_{\theta}(t=-\infty)=\sigma_{1}$ and $\sigma_{\theta}^{\perp}(t=-\infty)=\sigma_{3}$. Since at that time the dipole is prepared in the ground state of the Hamiltonian, we also have $\phi_{g}(t=-\infty)=\mid \sigma_{1}=-1>$. We now assume that at $t=-\infty$ the Schr$\ddot{o}$dinger and Heisenberg representations coincide. It follows that also
$\sigma_{\theta}^{H}(t=-\infty)=\sigma_{1}$, $\sigma_{\theta}^{\perp\,H}(t=-\infty)=\sigma_{3}$, and the Heisenberg state (at all times) is given by
\begin{equation}
\phi^{H}=\phi_{g}(t=-\infty)=\mid \sigma_{1}=-1>
\,.\label{eq:IV:21}
\end{equation}
Now in the adiabatic limit the dipole is in the instantaneous ground state of the Hamiltonian throughout. In the Heisenberg representation this reads $H_{1}^{H}(t)\phi_{g}(-\infty)=E(t)\sigma_{\theta}^{H}(t)\phi_{g}(-\infty)=-E(t)\phi_{g}(-\infty)$, (since at the adiabatic limit the Hamiltonian in the Heisenberg representation equals
~\cite{Heisenberg:Hamiltonian} $E(t)\sigma_{\theta}^{H}(t)$).  It then follows from $(\!\!~\ref{eq:IV:21})$ that
\begin{equation}
\sigma_{\theta}^{H}(t)\equiv\sigma_{1}
\,,\label{eq:IV:22}
\end{equation}
i.e., $\sigma_{\theta}^{H}$ is a constant of the motion. Hence $\sigma_{\theta}^{\perp\,H}(t)$ rotates in the perpendicular $y-z$ plane:
\begin{equation}
\sigma_{\theta}^{\perp\,H}(t)=\sigma_{3}\cos\gamma(t)+\sigma_{2}\sin\gamma(t)
\,,\label{eq:IV:23}
\end{equation}
where
\begin{equation}
\gamma(t)=\int_{-\infty}^{t}\omega(t')dt'
\,,\label{eq:IV:24}
\end{equation}
is the azimuthal angle. Equations $(\!\!~\ref{eq:IV:23})$ and $(\!\!~\ref{eq:IV:24})$  conform with the boundary condition $\sigma_{\theta}^{\perp\,H}(t=-\infty)=\sigma_{3}$. Substituting
$(\!\!~\ref{eq:IV:22})$ and $(\!\!~\ref{eq:IV:23})$ into $(\!\!~\ref{eq:IV:20})$ we finally obtain
\begin{equation}
\sigma_{3}^{H}(t)=\sin\theta(t)\sigma_{1}+\cos\theta(t)[\sigma_{3}\cos\gamma(t)+\sigma_{2}\sin\gamma(t)] \,.\label{eq:IV:25}
\end{equation}
\newline\newline\indent
 We next calculate the uncertainty of the relative phase$(\!\!~\ref{eq:IV:12})$. Using equations $(\!\!~\ref{eq:IV:13})$ and $(\!\!~\ref{eq:IV:25})$. We recall that since in the Heisenberg representations the dipole is in an eigenstate of $\sigma_{1}$, $\phi_{H}=\mid\sigma_{1}=-1>$, only the terms in $\sigma_{2}$ and $\sigma_{3}$ in $(\!\!~\ref{eq:IV:25})$ have a non vanishing contribution. Thus $\Delta\varphi_{rel}^{H}(t)=I_{2}^{2}+I_{3}^{2}$ where
\begin{eqnarray}
I_{2}(t)=-\int_{z(1)}^{z(2)}dz\frac{\partial f}{\partial z}\int_{-\infty}^{t}dt'g(t')\cos\theta (t')\cos\gamma (t')\equiv -\int_{z(1)}^{z(2)}dz I_{2}'(z,t)
\,,\label{eq:26}\\
I_{3}(t)=-\int_{z(1)}^{z(2)}dz\frac{\partial f}{\partial z}\int_{-\infty}^{t}dt'g(t')\cos\theta (t')\sin\gamma (t')\equiv -\int_{z(1)}^{z(2)}dz I_{3}'(z,t)
\,.\label{eq:IV:27}
\end{eqnarray}
To prove our claims no. 1 and no. 2 above, it suffices to calculate the integrals $I_{2}'$ and $I_{3}'$. Rather than $I_{2}'$, $I_{3}'$, we use
\begin{equation}
I'_{\pm}(z,t)=\int_{-\infty}^{t}dt'g(t')\cos\theta (t')e^{\pm i\gamma(t')}
\,.\label{eq:IV:28}
\end{equation}
For simplicity we first consider the same integrand without the $\cos\theta(t)$ factor.
We claim that for a general adiabatic function $g(t)\sim e^{-\varepsilon|t|}$
\begin{equation}
\tilde{I}'_{\pm}(z,t)=\int_{-\infty}^{t}dt'g(t')e^{\pm i\gamma(t')}=
\frac{g(t)e^{\pm i\gamma(t)}}{\pm i\omega(t)}-
\frac{g(-\infty)e^{\pm i\gamma(-\infty)}}{\pm i\omega(-\infty)}
\,,\label{eq:IV:29}
\end{equation}
where $\omega(t)=\frac{\partial\gamma}{\partial t}$. This can be readily proven by taking the derivatives of both sides of equation $(\!\!~\ref{eq:IV:29})$~\cite{derivative:1}.
Returning now to the integral of interest $(\!\!~\ref{eq:IV:28})$ we note that $(\!\!~\ref{eq:IV:15},\!\!~\ref{eq:IV:17},\!\!~\ref{eq:IV:18})$ $g'(t)\equiv g(t)\cos\theta(t) \sim\sin\theta(t)$
is itself adiabatic, since $ g(t)\cos \theta(t)\sim\frac{g(t)}{E(t)}
\sim \frac{e^{-\varepsilon|t|}}{\sqrt{1-e^{-2\varepsilon|t|}}}\approx  e^{-\varepsilon|t|}(1+\frac{1}{2} e^{-2\varepsilon|t|})\approx  e^{-\varepsilon|t|}$,
\\ $\alpha\sim f(z)\sim1$. Thus, we can apply $(\!\!~\ref{eq:IV:29})$ to
$g'(t)$, obtaining
\begin{eqnarray}
I'_{\pm}(z,t)&=&\int_{-\infty}^{t}dt'g(t')\cos\theta (t')e^{\pm i\gamma(t')}\nonumber\\
&=&\frac{g(t)\cos\theta (t)e^{\pm i\gamma(t)}}{\pm i\omega(t)}-
\frac{g(-\infty)\cos\theta (-\infty)e^{\pm i\gamma(-\infty)}}{\pm i\omega(-\infty)}
\,.\label{eq:IV:30}
\end{eqnarray}
All the above applies for $t$ in the regions $-\infty\ll t\ll t_{1}$ and $t_{2}\ll t\ll\infty$ where $g(t)$ varies, and trivially also in the region $t_{1}\ll t\ll t_{2}$ where $g(t)$ is constant.
\newline We next return to claim no. 1 above. To prove it, it suffices to show that $I'_{\pm}(z,\infty)=\int_{-\infty}^{\infty}dt g(t)\cos\theta (t)e^{\pm i\gamma(t)}=0$. From $(\!\!~\ref{eq:IV:30})$ we find that
\begin{equation}
I'_{\pm}(z,\infty)=\frac{g(\infty)\cos\theta (\infty)e^{\pm i\gamma(\infty)}}{\pm i\omega(\infty)}-
\frac{g(-\infty)\cos\theta (-\infty)e^{\pm i\gamma(-\infty)}}{\pm i\omega(-\infty)}
\,.\label{eq:IV:31}
\end{equation}
Both terms on the rhs vanish by the definition of $g(t)$. As for our claim no. 2, the non vanishing of  $\Delta\hat{\varphi}_{rel}(t)$ at finite times, it also trivially follows from $(\!\!~\ref{eq:IV:30})$.
\section{A concluding remark}
Our discussion so far has been largely conceptual. We now briefly address the feasibility of the experiments described above. In order to achieve an effective two-level system, a very small splitting between the ground state and the first excited state, as compared with all other energy differences of the system, is required.
\newline\newline\indent
In the present electric dipole context we model the "polarizable atom" by a single electron (or ion) which tunnels between two small conducting regions (or traps). The distance $d$ between these regions is much larger than their size $r$.  In this essentially double-well system the ground state (excited state) of the electron is the symmetric (antisymmetric) superposition $\frac{1}{\sqrt{2}}(\psi_{L}\pm\psi_{R})$, respectively, where $\psi_{L}$ $(\psi_{R})$ refers to the electron being in the ground state of the left (right) potential well. The splitting between the two levels ($2\alpha$ in our terminology) is thus rather small. Hence also our requirement that the interaction strength with the external particle be of order of $\alpha$ which determines the splitting can be readily met with no additional demand on the size of its charge.
\newline\newline\indent
To verify the applicability of the adiabatic approximation in the two-level system sketched above, we use the following qualitative argument. Let $r\approx$ O(100 ${\AA}$) and $d\approx$ O(1000 ${\AA}$). The choice of $r$ implies that the first excited state of the single electron in each of the circular patches separately is
\begin{equation}
\frac{\hbar^{2}}{2m_{e}r^{2}}\sim\frac{1}{100}eV\sim10^{13}sec^{-1}
\label{eq:33}
\end{equation}
above the ground state. On the other hand, the splitting 2$\alpha$ in the double well, which is proportional to the tunneling frequency $\omega_{t}$, is vastly smaller than $(\!\!~\ref{eq:33})$. This last frequency critically depends on the distance between the patches and can be tuned to a desired, sufficiently large, value. Specifically, we use this to ensure the adiabatic condition $\tau\omega_{t}\gg$1 where $\tau$ is the switching-on and -off time.\\
\appendix{\noindent \bf{Acknowledgement}}
This work has been supported in part by  the Israel Science Foundation  Grant No. 1125/10, and  The Wolfson Family Charitable Trust.\\
\appendix{\noindent \bf{Appendix A -- Proof of equation $(\!\!~\ref{eq:II:4})$}}
\newline\newline
With the Hamiltonian given by equation $(\!\!~\ref{eq:II:2})$ and the state of the system by $(\!\!~\ref{eq:II:3})$, multiplying the
    Schr$\ddot{o}$dinger equation  by $< \psi_{1}\mid $ or $< \psi_{2}\mid $, we obtain, respectively,
\begin{equation}
i\frac{\partial}{\partial t}\mid \phi _{1}>=[\frac{p_{l}^2}{2m}+\alpha\delta(x_{l})]\mid \phi _{1}>
\label{eq:em:1}
\end{equation}
and
\begin{equation}
i\frac{\partial}{\partial t}\mid\phi _{2}>=[\frac{p_{l}^2}{2m}+\alpha\delta(x_{l}-L)]\mid\phi _{2}>
\label{eq:em:2}
\end{equation}
Jointly, equations $(\!\!~\ref{eq:em:1})$ and $(\!\!~\ref{eq:em:2})$ are equivalent to equation $(\!\!~\ref{eq:II:4})$. These equations follow since the $\psi$'s do not depend on time, \mbox{$< \psi_{1}\mid \alpha\delta(x_{l}-x_{h}) \mid\psi_{1} >=\alpha\delta(x_{l})$},\\ \mbox{$<
\psi_{2}\mid \alpha\delta(x_{l}-x_{h}) \mid\psi_{2}>=\alpha\delta(x_{l}-L)$},\\ and \mbox{$< \psi_{i}\mid \alpha\delta(x_{l}-x_{h})
\mid\psi_{j} >=0$ for $i\neq j$}, since $\psi_{1}$, $\psi_{2}$, are orthogonal at all times since there is no recoil.
\newline\appendix{\noindent \bf{Appendix B -- Reflection and transmission from $\alpha\delta(x)$, $\alpha\delta(x-L)$}}
\newline\newline
The energy eigenstates of the Hamiltonian $H=\frac{p^2}{2m}+\alpha\delta (x) $
for $x<0$, $x>0$, are, respectively,
\begin{eqnarray}
\phi_{L}&=&e^{ipx}+Ae^{-ipx}\nonumber\\
\phi_{R}&=&Be^{ipx}
\,.\label{eq:rt:1}
\end{eqnarray}
Boundary conditions:\\
a.\begin{equation}
\phi_{L}(0)= \phi_{R}(0)\Rightarrow 1+A=B\,.
\label{eq:rt:2}
\end{equation}
b. Writing the stationary state Schr$\ddot{o}$dinger equation
\begin{equation}
-\frac{\hbar^{2}}{2m}\frac{\partial}{\partial x}\left(\frac{\partial\phi}{\partial x}\right)=[E-\alpha\delta(x)]\phi \,,
\label{eq:rt:3}
\end{equation}
Integrating from $-\Delta$ to $+\Delta$, letting $\Delta\rightarrow 0$,
\begin{equation}
\left[\frac{\partial\phi_{R}}{\partial x}\right|_{x=0}-\left[\frac{\partial\phi_{L}}{\partial x}\right|_{x=0}=\tilde{\alpha}\phi(0)
\,,\label{eq:rt:4}
\end{equation}
where the dimensionless $\tilde{\alpha}\equiv\frac{2m \alpha}{\hbar^{2}}$. Equations $(\!\!~\ref{eq:rt:1})$,  $(\!\!~\ref{eq:rt:2})$ and $(\!\!~\ref{eq:rt:4})$ yield
\begin{eqnarray}
A&=&\frac{-1}{1-\frac{2ip}{\tilde{\alpha}}}\approx -\left(1+ \frac{2ip}{\tilde{\alpha}}\right)
\rightarrow-1 \,\,\,\mbox{as }\tilde{\alpha}\rightarrow\infty\,,\nonumber\\
B&=&1+A \approx -\frac{2ip}{\tilde{\alpha}}\rightarrow 0 \,\,\,\mbox{as }\tilde{\alpha}\rightarrow\infty\,.
\label{eq:rt:5}
\end{eqnarray}
With the potential $\alpha\delta(x-L)$ and $\phi_{L}$, $\phi_{R}$,  now pertaining to $x<L$, $x>L$, respectively, the reflection and transmission coefficients at x=L are, respectively, $A'=e^{2ipL}A$
 and $B'=B$.
\newline\appendix{\noindent \bf{Appendix C -- Interference in the Heisenberg picture}}
\newline\newline
A concise and physically revealing way to calculate the relative phase presents itself in the Heisenberg picture. For a single particle in the superposition $\Psi=\frac{1}{\sqrt{2}}[\psi(x)+e^{i\varphi_{rel}}\psi(x-L)]$ the relative phase factor is given by the average the displacement operator
\begin{equation}
 \langle e^{ipL}\rangle=\langle\Psi(x)|e^{ipL}|\Psi(x)\rangle=\frac{e^{i\varphi_{rel}}}{2}\,,
\,\label{eq:displacement:1}
\end{equation}
having used $ e^{ipL}\psi(x)=\psi(x+L)$. The average displacement operator also equals the relevant Fourier transform of the particle's velocity distribution, since
\begin{equation}
 \langle e^{ipL}\rangle=\int P_{r}(p)e^{ipL}dp=F.T.[P_{r}(p)]\,.
\,\label{eq:displacement:2}
\end{equation}
Thus a change of the relative phase means that the velocity distribution has changed. Given the two particle state $(\!\!~\ref{eq:II:3})$ the average of the external particle's displacement operator equals $\langle\Psi(t)|e^{ip_{h}L}|\Psi(t)\rangle=\frac{1}{2}\langle\phi_{1}|\phi_{2}(t)\rangle=\frac{1}{2}\langle e^{i\varphi_{rel}}\rangle$ since in this case the relative phase factor is uncertain. For further discussion of this subject see ~\cite{aharonov:kaufherr}.
\newline\appendix{\noindent \bf{Appendix D -- The solution for the set-up of figure 4}}
\newline\newline
With an additional, infinite potential step at x=L$_{1}$, the Hamiltonian is
$H_{1}=H+V(x_{l}-L_{1})$, where $H$ is given by $(\!\!~\ref{eq:II:4})$ and
 $V(x_{l}-L_{1})=\infty$,  $x_{l}-L_{1}\geq 0$ , and $=0$ otherwise.
With the external particle in $\psi_{1}$, the internal particle sees the potential $\alpha\delta(x)+V(x-L_{1})$.  At the limit $\alpha\rightarrow\infty$ the $\delta$ function potential at the origin totally reflects the light particle. Thus energy eigenstate for $x<0$ is
\begin{equation}
\phi_{1}=\sin px.
\label{eq:ip:1}
\end{equation}
With the external particle in $\psi_{2}$, the internal particle sees the infinite potential step alone. The solution for $x\leq L_{1}$ is
\begin{equation}
\phi_{2}=\sin p(x-L_{1})\,,
\label{eq:ip:2}
\end{equation}
Thus the overall solution at all times t is given by
\begin{equation}
 \Psi(t)= c_{1}(t)(\psi_{1}+\psi_{2})e^{ipx}
+c_{2}(t)(\psi_{1}+e^{2ipL_{1}}\psi_{2})e^{-ipx}\,.
\label{eq:ip:3}
\end{equation}
\newline\appendix{\noindent \bf{Appendix E -- Equations $(\!\!~\ref{eq:II:14})$ and $(\!\!~\ref{eq:II:15})$}}
\newline\newline
We calculate here the interference term of the potential (i.e., the private potential) at x=L  and at x=0, and solve equation $(\!\!~\ref{eq:II:15})$. The following is based on Appendix B. $\tilde{\alpha}\equiv\frac{2m \alpha}{\hbar^{2}}$, $\hbar\equiv1$ thus $\tilde{\alpha}=2m \alpha$. At x=L we have
\begin{eqnarray}
\phi_{1}(x_{l}=L)&\approx&\frac{1}{\sqrt{W}}Be^{ip_{0} L}\approx\frac{1}{\sqrt{W}}(-\frac{iv_{0} }{\alpha})e^{ip_{0} L}\,,\nonumber\\
\phi_{2}(x_{l}=L)&\approx&\frac{1}{\sqrt{W}}B_{L}e^{ip_{0} L}=\frac{1}{\sqrt{W}}Be^{ip_{0} L}=\frac{1}{\sqrt{W}}(-\frac{iv_{0} }{\alpha})e^{ip_{0} L}\,,
\label{eq:appE:1}
\end{eqnarray}
\begin{equation}
<\phi_{1}|\alpha\delta(x_{l}-L)|\phi_{2}>=\alpha\phi_{1}^{\ast}(x_{l}=L)\phi_{2}(x_{l}=L)\approx\frac{v_{0}^{2}}{W\alpha}\rightarrow 0(\alpha\rightarrow\infty)\,.
\label{eq:appE:2}
\end{equation}
Thus the private potential vanishes at x=L for $\alpha\rightarrow\infty$. At x=0 we have
\begin{eqnarray}
  \phi_{1}(x_{l}=0)&\approx&\frac{1}{\sqrt{W}}B=\frac{1}{\sqrt{W}}(-\frac{2ip_{0} }{\tilde{\alpha}})
=\frac{1}{\sqrt{W}}(-\frac{ip_{0} }{m\alpha})\,,\nonumber\\
\phi_{2}(x_{l}=0)&\approx&\frac{1}{\sqrt{W}}\left(1- e^{2ip_{0} L}\right)\,,
\label{eq:appE:3}
\end{eqnarray}
\begin{eqnarray}
<\phi_{1}|\alpha\delta(x_{l})|\phi_{2}>&=&
\alpha\phi_{1}^{*}(x_{l}=0)\phi_{2}(x_{l}=0)\nonumber\\
&=&\alpha\frac{1}{\sqrt{W}}(\frac{ip_{0}}{m\alpha})\frac{1}{\sqrt{W}}\left(1- e^{2ip_{0} L}\right)
=i\frac{v_{0}}{W}\left(1- e^{2ip_{0} L}\right).
\label{eq:appE:4}
\end{eqnarray}
Note that the private potential at x=0 is finite. Using $(\!\!~\ref{eq:appE:2})$ and $(\!\!~\ref{eq:appE:4})$ we obtain from equation $(\!\!~\ref{eq:II:14})$
\begin{equation}
\frac{d}{dt}\left<e^{i\varphi_{rel}}\right>=-\frac{v_{0}}{W}\left(1- e^{2ip_{0} L}\right)\,.
\label{eq:appE:5}
\end{equation}
This is equation $(\!\!~\ref{eq:II:15})$. Its solution is
\begin{equation}
\left<e^{i\varphi_{rel}}(t)\right>=\left<e^{i\varphi_{rel}}(0)\right>
-\frac{v_{0}t}{W}+ \frac{v_{0}t}{W}e^{2ip_{0}L}=\frac{W-v_{0}t}{W}+ \frac{v_{0}t}{W}e^{2ip_{0}L}\,.
\label{eq:appE:6}
\end{equation}
\newline\appendix{\noindent \bf{Appendix F -- The average momentum equals the gradient of the phase for a very narrow stationary Gaussian wave packet}}
\newline\newline
The heavy, i.e., non recoiling external particle may be described by a stationary Gaussian wave packet. With the external particle located at $0\leq x\leq L$ we have
\begin{equation}
\psi(x_{h}-x)=ne^{-\frac{(x_{h}-x)^{2}}{2\Delta^{2}}}e^{i\varphi(x_{h})}\equiv ne^{-f(x_{h}-x)}e^{i\varphi (x_{h})}\,,
\label{eq:sg:1}
\end{equation}
such that $\int\psi^{*}\psi dx_{h} =1$. $\langle p_{h}\rangle=\int\psi^{*}\frac{1}{i}\frac{\partial\psi}{\partial x_{h}} dx_{h}\,.$ The imaginary part of $\langle p_{h}\rangle$ vanishes since $f$ is an even function of $x_{h}$ and $\frac{\partial f}{\partial x_{h}}$ is odd. The real part equals $\langle p\rangle_{h}=\frac{\partial \varphi}{\partial x_{h}}(x)$ since the very narrow stationary Gaussian distribution
$n^{2}e^{-2f(x_{h}-x)}$ behaves like a $\delta$ function at $x$. Thus, in this case, the average momentum equals the gradient of the phase.
\newline\appendix{\noindent \bf{Appendix G -- The uncertainty of the relative phase}}
\newline\newline
First note that for an ensemble of N particles all in the same state $\psi_{1}(n)+e^{i\varphi(x,t)}\psi_{2}(n)$ one can define an Hermitean operator $\hat{\varphi}$ whose eigenvalue is $\varphi(x,t)$. From $\delta\varphi_{rel}(t)=\int_{0}^{L}\int_{0}^{t}\langle F\rangle dt' dx =\langle \int_{0}^{L}\int_{0}^{t} F dt' dx\rangle$ we infer that $\delta\hat{\varphi}_{rel}(t)=\int_{0}^{L}\int_{0}^{t}\hat{F} dt' dx=\int_{0}^{L}\hat{I}(x,t)dx$  where $\hat{I}$ is the impulse of the force. Its uncertainty is given by
\begin{equation}
\Delta \delta\hat{\varphi}_{rel}(t)=\Delta\int_{0}^{L} \hat{I}(x,t)dx\leq\int_{0}^{L}\Delta \hat{I}(x,t)dx\leq\int_{0}^{L}\int_{0}^{t}\Delta\hat{F} dt' dx\,,
\label{eq:uncertainty:1}
\end{equation}
having used Schwartz's inequality. Thus a necessary condition for preserving the coherence of the entangled external particle is given by
\begin{equation}
\int_{0}^{L}\int_{0}^{t}\Delta\hat{F} dt' dx< 2\pi\,.
\label{eq:uncertainty:2}
\end{equation}
In other words, the locally measurable forces and their impulses provide only a necessary condition for coherence of an external particle that interacts with the environment. In particular it follows from equation $(\!\!~\ref{eq:uncertainty:1})$, and from $\Delta \delta\hat{\varphi}_{rel}(t)\geq 0$, that both initially and finally the uncertainty of the change of the relative phase vanishes, i.e., $\Delta \delta\hat{\varphi}_{rel}(0)=\Delta \delta\hat{\varphi}_{rel}(T)=0$. At these times there is no reflection and therefore $\Delta\hat{F}=0$. Thus only at intermediate times the change of the relative phase is uncertain.
Note also that the impulse of the force $ I(x,t)=\int_{0}^{t}F dt'=p_{t}-p_{0}=\delta p_{t}$ is determined by measuring directly the change of the particle's momentum, rather than by measuring the initial and final momenta separately.
\newline\appendix{\noindent \bf{Appendix H -- Derivation of equation $(\!\!~\ref{eq:III:3})$}}
\newline\newline
 To calculate the polarization, let $\alpha\equiv \lambda_{1}$, $g(t)f(z)\equiv \lambda_{2}$. The stationary Schr$\ddot{o}$dinger equation for the
    ground state
    is
\[\left( \begin {array}{rr}\lambda_{2}& \lambda_{1}\\\lambda_{1}&-\lambda_{2} \end{array}   \right)
\left(\begin {array}{c}\cos \theta\\\sin\theta   \end{array}\right)=-|E|\left(\begin {array}{c}\cos \theta\\\sin\theta
\end{array}\right)\,.\] We obtain
$\langle\sigma_{3}\rangle=\cos^{2}\theta-\sin^{2}\theta=-\frac{\lambda_{2}}{|E|}
=-\frac{\lambda_{2}}{\sqrt{\lambda^{2}_{1}+\lambda^{2}_{2}}}$. For the excited state $\langle\sigma_{3}\rangle=\frac{\lambda_{2}}{|E|}$.
\newline\appendix{\noindent \bf{Appendix I -- Proof of equation $(\!\!~\ref{eq:III:6})$}}
\newline\newline
Writing the dipole's Hamiltonian $H_{1}(z)=H_{F}+V(z,t)$, let $z=z_{0}$ and consider a virtual displacement $z_{0} \rightarrow z_{0}+\Delta z$ of the external particle. We use stationary state perturbation theory to calculate the corresponding change in the ground state energy of the dipole. We have at $z=z_{0}$
\begin{equation}
H_{1}(z_{0})\mid\phi_{g}(z_{0})>=E_{g}(z_{0})\mid\phi_{g}(z_{0})>
\label{eq:boE:1}
\end{equation}
and at $z=z_{0}+\Delta z$, to first order in $\Delta z$
\begin{equation}
H_{1}(z_{0}+\Delta z)\approx H_{F}+V(z_{0})+\frac{\partial V}{\partial z}(z_{0})\Delta z\equiv H_{1}(z_{0})+\Delta V
\label{eq:boE:2}
\end{equation}
\begin{equation}
\left[ H_{1}(z_{0})+\Delta V \right]\mid\phi_{g}(z_{0})+\delta\phi_{g}>= \left[ E_{g}(z_{0})+\Delta E_{g}\right]\mid \phi_{g}(z_{0})+\delta\phi_{g}>
\label{eq:boE:3}
\end{equation}
Subtracting $(\!\!~\ref{eq:boE:1})$ from $(\!\!~\ref{eq:boE:3})$, neglecting second order terms in $\Delta$ and $\delta$, we obtain
\begin{equation}
\Delta E_{g}=<\phi_{g} \mid \Delta V \mid \phi_{g}>=\left\langle \frac{\partial V}{\partial z}\right\rangle \Delta z
=-\langle F\rangle \Delta z
\label{eq:boE:4}\,.
\end{equation}
Renaming $z_{0}+\Delta z\equiv z_{0}$, $H_{0}+\Delta V\equiv H_{0}$ and repeating the above, we finally obtain equation $(\!\!~\ref{eq:III:6})$ after a succession of such steps and summation.
\newline\appendix{\noindent \bf{Appendix J -- The Born Oppenheimer potential of the test particle}}
\newline\newline
The Born Oppenheimer potential experienced by the test particle is the ground state energy of the Hamiltonian $H_{1}'$. See equations $(\!\!~\ref{eq:III:8})$ and $(\!\!~\ref{eq:III:9})$. Assuming for simplicity $g(t)\equiv 1$ we have
\begin{eqnarray}
(E')^{2}&=&\alpha^{2}+\left[ f(z_{0})+\varepsilon f(z_{t})  \right]^{2}\nonumber\\
        &\approx& \alpha^{2}+f^{2}(z_{0})+2\varepsilon f(z_{0})f(z_{t})  \nonumber\\
        &=&\left[\alpha^{2}+f^{2}(z_{0}) \right]\left[1+ \frac{2\varepsilon f(z_{0})f(z_{t})}{\alpha^{2}+f^{2}(z_{0})} \right]
\,.\label{eq:boT:1}
\end{eqnarray}
Taking the negative square root of $(\!\!~\ref{eq:boT:1})$ and  using $(\!\!~\ref{eq:III:3})$ we obtain to first order in $\varepsilon$ , $E'_{g}=E_{g}(z_{0})+\varepsilon[\langle\sigma_{3}\rangle(z_{0})]f(z_{t})$.
\newline\appendix{\noindent \bf{Appendix K -- The non vanishing private potential in vacuum}}
\newline\newline
We here calculate the condition for the vanishing everywhere of the public potential and the forces, during $t_{1}\leq t\leq t_{2}$. We show that under this condition the private potential difference between the Faraday cages is, in general, different from zero.
The distant wave packet enclosed in the grounded Faraday cage does not contribute to the  polarization.
Denoting $\epsilon_{i}=\sqrt{\alpha_{i}^{2}+\beta_{i}^{2}g^{2}(0)f^{2}[z(2)]}=\epsilon_{i}(t_{1}\leq t\leq t_{2} )$, the condition for the vanishing of the public potential during $t_{1}\leq t\leq t_{2}$ is
\[\frac{\beta_{1}^{2}}{\epsilon_{1}}-\frac{\beta_{2}^{2}}{\epsilon_{2}}=0\;
\Rightarrow \beta_{1}^{2}\epsilon_{2}- \beta_{2}^{2}\epsilon_{1}=0\,.\]
In general, it is satisfied at a single point $\tilde{z}(2)$. On the other hand, the Born Oppenheimer or private potential is
$\mbox{$U_{BO}(z,t)=E_{g}^{1}(z,t)+E_{exc}^{2}(z,t)$}$. Thus, during $t_{1}\leq t\leq t_{2}$,  $U_{BO}[\tilde{z}(2),t]-U_{BO}[z(1),t]=\epsilon_{2}-\alpha_{2}-[\epsilon_{1}-\alpha_{1}]$
which in general is different from zero.


\begin{thebibliography}{999}
\bibitem{derived:explained}
 Specifically, equation $(\!\!~\ref{eq:II:9})$ is due to reference ~\cite{stern:aharonov}. It is derived and explained below.
\bibitem{finite:plates}
However, the two wave packets can be positioned on the two respective sides of the (as yet closed) condenser only if the condenser plates are finite in at least one of the $y$ and $z$ directions. We therefore use a parallel-plate condenser with very large but finite plates as an approximation. When the condenser is opened, the two wave packets are located very far from its edge.
\bibitem{narrow:tunnel}
The width of the tunnel d is small enough so that the lowest excitation in the transverse direction $\varepsilon_{transverse}=\frac{\hbar^{2}}{2md^{2}}$ is much larger than all other energy scales in the problem, and therefore the motion can be viewed as one-dimensional.
\bibitem{paradox:one}
 We have here implicitly, and erroneously, assumed the potentials of a test particle, see below.
\bibitem{complex:conjugate}
The complex conjugate of equation  $(\!\!~\ref{eq:em:1})$ is
\begin{eqnarray}
-i\frac{\partial}{\partial t}<\phi _{1}\mid=<\phi _{1}\mid[\frac{p_{l}^2}{2m}+\alpha\delta(x_{l})]\,.
\label{eq:cc:1}
\end{eqnarray}
\bibitem{Heisenberg:Hamiltonian}
Strictly speaking, the Hamiltonian in the Heisenberg representation is given by
$H^{H}=U^{\dag}HU+iU^{\dag}\frac{\partial U}{\partial t}$. It turns out that in the adiabatic limit the second term can be neglected.
\bibitem{derivative:1}
The derivative of the integral is, by definition, $g(t)e^{\pm i\gamma(t)}$. The derivative of the rhs yields the desired  $g(t)e^{\pm i\gamma(t)}$ and in addition\\
    \[\left(\frac{\varepsilon e^{\varepsilon t}(\pm i\omega(t))-i\frac{\partial\omega}{\partial t}e^{\varepsilon t} }{i\omega^{2}(t)}\right)e^{\pm i\gamma(t)}\]
    When $\varepsilon\rightarrow o$ the first term vanishes. Also, in the adiabatic limit, assumed throughout, the fractional change of the instantaneous energy or frequency $\omega(t)$, over the period $\tau(t)\sim\frac{1}{\omega(t)}$, the turning -on and -off time, is small. This allows us to neglect the second term as well.
\end{thebibliography}
\end{document}